\documentclass[twocolumn]{aastex63}
\usepackage{amsmath}

\newcommand{\casa}{{\sc casa}}
\newcommand{\sfrsd}{$\Sigma_{\mathrm{SFR}}$}
\newcommand{\ssd}{$\Sigma_{*}$}
\newcommand{\gsd}{$\Sigma_{\mathrm{mol}}$}
\newcommand{\aco}{$\alpha_{\mathrm{CO}}$}
\newcommand{\smol}{$\sigma_{\mathrm{mol}}$}
\newcommand{\tdep}{$t_{\mathrm{dep}}$}
\newcommand{\torb}{$t_{\mathrm{orb}}$}
\newcommand{\jupone}{CO(1$-$0)}
\newcommand{\juptwo}{CO(2$-$1)}
\newcommand{\jupthree}{CO(3$-$2)}
\newcommand{\jupfour}{CO(4$-$3)}

\newcommand{\CII}{\mbox{\rm [C{\small II}]}}
\newcommand{\CI}{\mbox{\rm [C{\small I}]}}
\newcommand{\Htwo}{\mbox{\rm H$_{2}$}}

\newcommand{\HII}{\mbox{\rm H{\small II}}}
\newcommand{\kms}{km\,s$^{-1}$}

\received{}
\revised{}
\accepted{}
\submitjournal{}

\shorttitle{Disk Turbulence and Star Formation Regulation}
\shortauthors{Lenki\'{c} et al.}
\graphicspath{{./}{Figures/}}

\begin{document}

\title{Disk Turbulence and Star Formation Regulation in High$-z$ Main Sequence Analogue Galaxies}

\correspondingauthor{Laura Lenki\'{c}}
\email{laura.lenkic@gmail.com}

\author[0000-0003-4023-8657]{Laura Lenki\'{c}}
\affiliation{Department of Astronomy, University of Maryland, College Park, MD 20742, USA}
\affiliation{SOFIA Science Center, USRA, NASA Ames Research Center, M.S. N232-12, Moffett Field, CA 94035, USA}
\affiliation{Jet Propulsion Laboratory, California Institute of Technology, 4800 Oak Grove Dr., Pasadena, CA 91109, USA}

\author[0000-0003-0645-5260]{Deanne B. Fisher}
\affiliation{Centre for Astrophysics and Supercomputing, Swinburne University of Technology, PO Box 218, Hawthorn, VIC 3122, Australia}
\affiliation{ARC Centre of Excellence for All Sky Astrophysics in 3 Dimensions (ASTRO 3D)}

\author[0000-0002-5480-5686]{Alberto D. Bolatto}
\affiliation{Department of Astronomy, University of Maryland, College Park, MD 20742, USA}

\author[0000-0003-1774-3436]{Peter J. Teuben}
\affiliation{Department of Astronomy, University of Maryland, College Park, MD 20742, USA}

\author[0000-0003-2508-2586]{Rebecca C. Levy}
\altaffiliation{NSF Astronomy and Astrophysics Postdoctoral Fellow}
\affiliation{Steward Observatory, University of Arizona, Tucson, AZ 85721, USA}

\author[0000-0003-0378-4667]{Jiayi Sun}
\altaffiliation{NASA Hubble Fellow}
\affiliation{Department of Astrophysical Sciences, Princeton University, 4 Ivy Lane, Princeton, NJ 08544, USA}

\author[0000-0002-2775-0595]{Rodrigo Herrera-Camus}
\affiliation{Departamento de Astronomía, Universidad de Concepción, Barrio Universitario, Concepción, Chile}

\author[0000-0002-3254-9044]{Karl Glazebrook}
\affiliation{Centre for Astrophysics and Supercomputing, Swinburne University of Technology, PO Box 218, Hawthorn, VIC 3122, Australia}
\affiliation{ARC Centre of Excellence for All Sky Astrophysics in 3 Dimensions (ASTRO 3D), Australia}

\author[0000-0002-1527-0762]{Danail Obreschkow}
\affiliation{International Centre for Radio Astronomy Research (ICRAR), University of Western Australia, Crawley, WA 6009, Australia}

\author[0000-0002-4542-921X]{Roberto Abraham}
\affiliation{Department of Astronomy \& Astrophysics, University of Toronto, 50 St. George Street, Toronto, ON M5S 3H4, Canada}



\begin{abstract}
The gas-phase velocity dispersions in disk galaxies, which trace turbulence in the interstellar medium, are observed to increase with lookback time. However, the mechanisms that set this rise in turbulence are observationally poorly constrained. To address this, we combine kiloparsec-scale ALMA observations of \jupthree{} and \jupfour{} with HST observations of H$\alpha$ to characterize the molecular gas and star formation properties of seven local analogues of main sequence galaxies at $z \sim 1-2$, drawn from the DYNAMO sample. Investigating the ``molecular gas main sequence'' on kpc-scales, we find that galaxies in our sample are more gas-rich than local star-forming galaxies at all disk positions. We measure beam smearing corrected molecular gas velocity dispersions and relate them to the molecular gas and star formation rate surface densities. Despite being relatively nearby ($z \sim 0.1$), DYNAMO galaxies exhibit high velocity dispersions and gas and star formation rate surface densities throughout their disks, when compared to local star forming samples. Comparing these measurements to predictions from star formation theory, we find very good agreements with the latest feedback-regulated star formation models. However, we find that theories which combine gravitational energy dissipation from radial gas transport with feedback over-estimate the observed molecular gas velocity dispersions.
\end{abstract}



\section{Introduction} \label{sec:intro}
A key result of large numbers of surveys over the past decade is the rise in gas-phase velocity dispersions from the local Universe to those galaxies at lookback times of $\sim$10 billion years \citep[$z \sim 2$; e.g.][]{Law2009,genzel11,forsterschreiber11,Epinat2012,Kassin2012,Wisnioski2015,Ubler2019}. This velocity dispersion evolution is correlated with increases in many physical properties of galaxies, especially those related to star formation and the interstellar medium (ISM). Galaxies of a fixed stellar mass show increases in molecular gas surface density, star formation rate surface density, and gas fraction with lookback time \citep[see reviews by][]{Glazebrook2013,forsterschreiber2020,Tacconi2020}. They also become more compact and their morphologies become dominated by patches of high star formation, typically called clumps. Understanding what mechanisms set the rise in velocity dispersion in disk galaxies, and how it is linked to other changes in the ISM, is therefore needed to build accurate models of galaxy evolution. A historical challenge to this has been the lack of resolved observations of the cold gas velocity dispersions in galaxies with higher gas fractions ($f_\mathrm{gas}>15$\%) and clumpy morphologies.

The observed velocity dispersion in disks is typically interpreted to reflect the turbulence of gas in the ISM. Turbulence is a key mechanism to help galaxies regulate their star formation \citep{bournaud10}. In many theories, the gas collapse induced by the gravitational potential well of gas, stars, and dark matter is balanced by the thermal, turbulent, radiation, and magnetic pressures, where turbulence is often invoked as the primary balancing force. An important question in understanding this equilibrium state is determining what mechanisms predominantly drive turbulence in the ISM. Under the theory of self-regulated star formation, momentum injected by stellar feedback is enough to drive the turbulence needed to balance the vertical weight of the ISM \citep{shetty12,faucher-giguere13}. Alternatively, theories of gas dynamics predict that the accretion powered release of gravitational potential energy via radial inflows of gas through a galactic disk are the primary drivers of turbulence \citep{krumholz16} or that both stellar feedback and gas transport are required to explain observations of the velocity dispersion--star formation rate surface density relation \citep{elmegreen10,klessen10,krumholz18}.

An important, recent advancement in this area is the increasing number of observations of velocity dispersion in cold gas tracers. Recent work finds that CO-based measurements of the velocity dispersion are systematically lower than those using ionized gas tracers \citep{levy18,Girard2019,Ubler2019,girard21,Liu2023}. Typical differences between the measured velocity dispersion in ions and molecules can be quite high, $\sigma_\mathrm{ion}-\sigma_\mathrm{mol}\sim 30-50$~km~s$^{-1}$. \cite{girard21} compiled a sample of galaxies and showed that the offset is a roughly constant ratio, $\sigma_\mathrm{ion}/\sigma_\mathrm{mol}\sim2-3$, with respect to gas fraction and redshift, while \cite{Ubler2019} argued that the offset may evolve, becoming smaller at higher redshift. However, there are not sufficient data points at higher redshift to conclusively determine if $\sigma_\mathrm{ion}/\sigma_\mathrm{mol}$ evolves with redshift or not. Nevertheless, there appears to be agreement overall that this offset exists and is significantly larger than a simple correction for the thermal broadening of \HII{} regions, which is rarely more than 15~km~s$^{-1}$ \citep{krumholz16}.

Molecular gas represents a significantly larger fraction of the ISM mass than ions, and thus may be more representative of turbulence in galaxies, especially the part that is involved in regulating star formation. \cite{girard21} showed that this difference has important implications for comparison to theory, and found that models of gravity+feedback overestimate the velocity dispersion when measured with CO. Recent simulations found similar results: Feedback acts to stratify the gas disk and generates differences in the velocity dispersion of ionised and molecular gas \citep{Ejdetjarn2022,Rathjen2023}.

Gas velocity dispersion is not the only means of studying the regulation of star formation in galaxies. The Kennicutt-Schmidt relation \citep[hereafter KS relation;][]{Schmidt1959,kennicutt98b,kennicutt12}, which ties the surface density of star formation (\sfrsd{}) to the surface density of gas (\gsd{}) through a power law relation with a slope of $N \sim 1.4$, is a widely studied relationship that is discussed in many of the theories described above. It is frequently used as a basic metric of how rapidly gas is consumed by star formation. The KS relation spans several orders-of-magnitude in both the star formation rate (SFR) and gas surface density, and it holds for both normal star-forming galaxies and starbursting systems \citep{kennicutt21}.

A strong relationship between \sfrsd{} and \gsd{} appears to persist on ${\sim} 1$~kpc scales \citep{leroy08,sanchez21}, though subtle differences in the slope of the relation may exist. A large number of authors studying local Universe spirals argue for slopes that are close to unity \citep{Bigiel2008,leroy13,Sun2023}. At $z \sim 1-2$ the picture is far less clear. Only a handful of galaxies have resolved observations of both \sfrsd{} and \gsd{}. As recently discussed in \cite{Fisher2022}, the combined sample of targets from the literature is quite heterogeneous, hindering our ability to derive any general conclusions about the nature of this relationship at high \sfrsd{}. Systematic studies of selected samples of galaxies measuring the resolved KS relation at $z>0.5$ remain absent from the literature \citep[see discussion in][]{Tacconi2020}.

In this paper, we study the molecular gas velocity dispersions (\smol), molecular gas surface densities (\gsd), and star formation rate surface densities (\sfrsd) of {\bf seven highly turbulent}, nearby ($z \sim 0.1$) galaxies from the DYnamics of Newly Assembled Massive Objects \citep[DYNAMO;][]{green14} sample. The gas fractions \citep{fisher14,White2017} and ionized gas velocity dispersions \citep{green14,oliva-altamirano18} of these galaxies are most consistent with main sequence star forming galaxies at $z \sim 1$. Moreover, the H$\alpha$ morphology is consistent with so-called ``clumpy" galaxies \citep{fisher17a,lenkic21,ambachew22}. Furthermore, the star formation rates and stellar masses of DYNAMO galaxies place them on the main sequence of star formation at $z \sim 2$ rather than the local ($z \sim 0.1$) one \citep{fisher19}. Their resemblance to high-redshift systems and proximity to us allows us to probe the turbulence powering mechanisms in gas-rich galaxies on kpc scales.

We combine H$\alpha$ observations from the Hubble Space Telescope with \jupthree{} and \jupfour{} observations from the Atacama Large Millimeter/sub-millimeter Array to study the \smol$-$\gsd{} and \smol$-$\sfrsd{} relations resolved on ${\sim} 1-2$~kpc scales, and compare these to results from simulations and expectations from star formation theory. This paper is structured as follows: \S\ref{sec:obs} describes our observations and data reduction, \S\ref{sec:methods} describes how we derive molecular gas surface density, star formation rate surface density, stellar mass surface density, and velocity dispersions, \S\ref{sec:results} presents our results on the ``molecular gas main sequence'', KS relation, and the \smol{}$-$\gsd{} and \smol{}$-$\sfrsd{} relations, \S\ref{sec:discussion} compares our results to expectations from theories of star formation regulation, and finally we conclude in \S\ref{sec:conclusion}.

Throughout this work, we assume $\Lambda$CDM cosmology with H$_{0} = 69.6$~km\,s$^{-1}$, $\Omega_{m} = 0.286$, and $\Omega_{\Lambda} = 0.714$, and a Kroupa initial mass function \citep{kroupa01}.

\begin{deluxetable*}{llcccccccc}
\tablecaption{Galaxy properties\label{tab:properties}}
\tablewidth{0pt}
\tablehead{
Galaxy & \colhead{z} & \colhead{M$_{*}$} & \colhead{SFR} & \colhead{$f_\mathrm{gas} \, ^a$} &  \colhead{$\sigma_{0,\mathrm{ion}} \, ^{b}$} & \colhead{$\sigma_{0,\mathrm{mol}} \, ^{b}$} & \colhead{$\sigma_{m,\mathrm{mol}} \, ^{c}$} & \colhead{$\Sigma_\mathrm{SFR} \, ^{a}$} & 
\colhead{CO Beam FWHM} \\
\colhead{} & \colhead{} & \colhead{[$10^{10}$~M$_{\odot}$]} & \colhead{[M$_{\odot}$\,yr$^{-1}$]} & \colhead{} &  \colhead{[\kms{}]}& \colhead{[\kms{}]} &  \colhead{[\kms{}]} & \colhead{[log\,(M$_{\odot}$\,yr$^{-1}$\,kpc$^{-2}$)]} & \colhead{[kpc (arcsec)]}
}
\startdata
C13-1 & 0.07876 & 3.58 & $5.06 \pm 0.5$ & $0.06 \pm 0.02$ &  26 & 8 & 17 & -1.64 $\pm$ 0.05 & 1.60 (1.07) \\
D13-5 & 0.07535 & 5.38 & $17.48 \pm 0.45$  & $0.36\pm0.02$ &  40 & 12 & 22 & -0.48 $\pm$  0.02 & 1.58 (1.10) \\
D15-3 & 0.06712 & 5.42 & $ 8.29 \pm 0.35$ &  $0.17\pm0.04$ &  25 & 8 & 14 & -0.87 $\pm$ 0.03 & 1.24 (0.96) \\
G04-1 & 0.12981 & 6.47 &  $ 21.32 \pm 1$ & $0.33\pm0.04$ &  30 & 13 & 16 & -0.65 $\pm$  0.03 & 0.98 (0.42) \\
G08-5 & 0.13217 & 1.73 &   $ 10.04 \pm 1$ & $0.30\pm0.05$ & 36 & 15 & 20 & -0.63 $\pm$ 0.04 & 0.95 (0.40) \\
G14-1 & 0.13233 & 2.23 &  $ 6.9 \pm 0.5$ & $0.77\pm0.08$ & 71 & 27 & 35 & -0.36 $\pm$  0.04 & 1.02 (0.43) \\
G20-2 & 0.14113 & 2.16 &  $18.24 \pm 0.35$ & $0.21\pm0.05$ & 36 & 9 & 23 & -0.48 $\pm$  0.02 & 3.08 (1.23) \\
\enddata
\tablecomments{\\
$^a$ \, Values from \citet{fisher17a}, \citet{White2017} and \citet{fisher19}. \\
$^b$ \, Values from \citet{girard21}. Uncertainty on the ionized gas velocity dispersion, $\sigma_{0,\mathrm{ion}}$, is $3-5$~\kms{}, and on the molecular gas velocity dispersion, $\sigma_{0,\mathrm{mol}}$, is $2-3$~\kms{}. \\
$^c$ \, The median velocity dispersion in this work determined by fitting each line-of-sight CO line profile with a Gaussian and correcting for beam smearing is given by $\sigma_{m,\mathrm{mol}}$.
}
\end{deluxetable*}  

\section{Observations} \label{sec:obs}
\subsection{Galaxy Sample} \label{subsec:sample}
The DYNAMO sample was first defined by \citet{green14}, who selected galaxies from the MPA-JHU Value Added Catalog of the Sloan Digital Sky Survey based on their redshifts and H$\alpha$ emissions. The sample comprises 67 galaxies, of which half have L$_{\mathrm{H}\alpha} > 10^{42}$ erg\,s$^{-1}$. A significant amount of previous work has compared DYNAMO to $z\sim1$ systems. DYNAMO galaxies, including those studied in this work, have been shown to be much more gas rich than local Universe galaxies, with gas fractions of order $10-70$\% \citep{fisher14,White2017,fisher19}. This is higher than local Universe main-sequence galaxies by roughly a factor of a few \citep{Saintonge2011}, and similar to galaxies at $z \sim 1$ \citep{Tacconi2020}. DYNAMO galaxies have likewise been shown to be clumpy in both ionized gas \citep{fisher17a} and $U$-band star light \citep{lenkic21,ambachew22}, where clumps are defined as done in CANDELS survey using the ratio of clump light to total galaxy light \citep{guo15}.

The kinematics of DYNAMO galaxies are similar to $z\sim1$ galaxies in more ways than simply the velocity dispersion. \cite{fisher17b} shows that DYNAMO galaxies are consistent with low values of Toomre~$Q$, $Q \sim0.5-1.5$ \citep[see also][]{White2017} and that the Toomre values correlate to the clump sizes. \cite{Obreschkow2015} show that DYNAMO galaxies are low angular momentum outliers to local relationships between specific angular momentum ($j$) and galaxy mass. They have $j/M_{*}$ that is more similar to observations at $z\sim1$. Overall, DYNAMO galaxies have been shown in many ways to be similar in properties to $z\sim1$ galaxies. 

In this work, we use seven targets from the intersection of the samples of \cite{fisher17a}, which provides HST imaging of DYNAMO targets and \cite{lenkic23}, which provides ALMA CO maps for the same galaxies. In Table~\ref{tab:properties}, we summarize the basic properties of the galaxies in this paper.

\subsection{ALMA and HST Observations} \label{subsec:obs}
The ALMA observations we use are associated with project codes 2017.1.00239.S (PI: D. B. Fisher) and 2019.1.00447.S (PI: R. Herrera-Camus). These observations were imaged using \texttt{tclean} with the Common Astronomy Software Application \citep[\casa{}][]{mcmullin07} with \texttt{weighting=``briggs''} and \texttt{robust=0.5}. Detailed information on the data processing and data products can be found in \citet{lenkic23}.

In addition to the ALMA observations of CO in our DYNAMO galaxies, we make use of HST observations of H$\alpha$ as a tracer of the star formation rate (PID
12977; P.I.: I. Damjanov), and HST WFC3/IR F125W observations ($\sim J$ band) as a tracer of the stellar mass (PID 15069; P.I.: D. B. Fisher). For details on the reduction and analysis of these observations, see \citet{fisher17a} and \citet{ambachew22}, respectively. 

To investigate the kpc-scale KS relation and relation between \smol{} and \gsd{} and \sfrsd{}, we match the pixel scale and resolution of the H$\alpha$ and F125W observations to that of the \jupthree{} where available, and \jupfour{} otherwise. To achieve this, we convolve the HST observations with a two-dimensional Gaussian function whose FWHM is equal to the circularized beam of the corresponding ALMA observation. Then, we re-project and re-grid them to match the WCS information and pixel scale of the CO observations using the {\sc Python astropy} package \texttt{reproject}\footnote{\href{https://reproject.readthedocs.io/en/stable/index.html}{https://reproject.readthedocs.io/en/stable/index.html}}, which assumes input images have surface brightness units. Our input images have count rate units, thus we scale the reprojected images by the squared ratio of the new pixel scale and old pixel scale to conserve flux.

\section{Methods} \label{sec:methods}
The properties we are interested in measuring and studying are: (1) the stellar mass surface densities (\ssd{}), (2) the molecular gas surface density (\gsd) adopting a constant and variable CO-to-H$_2$ conversion factor (\aco), (3) the SFR surface density (\sfrsd), and (4) the molecular gas velocity dispersions (\smol{}).

For each data set, we define a ``grid'' of circular, beam-sized apertures centered on the galaxy, and a second that is offset from the center by $0.5 \times$ the beam FWHM in both the $x$ and $y$ directions (RA and Dec.) to cover the gaps of the first grid.

\subsection{Stellar Mass Surface Density}
We measure the stellar mass surface density (\ssd{}) from HST F125W observations, matched to the resolution and pixel scale of the CO observations. We perform aperture photometry along every beam-sized line-of-sight in the two grids, as described above. In addition, we perform aperture photometry in the same way on HST WFC3/UVIS F336W observations and then measure the $F336W-F125W$ color. \citet{ambachew22} studied the stellar masses of clumps in a sample of DYNAMO galaxies and derived mass-to-light ratios based on HST colors. Therefore, we use the $F336W-F125W$ colors we measure to derive a mass-to-light ratio ($\Upsilon_{*,\mathrm{F125W}}$) for each line-of-sight beam-sized aperture from the relation:

\begin{equation}
    \mathrm{log} \, \Upsilon_{*,\mathrm{F125W}} = 0.195 \times (F336W - F125W) - 1.187
\end{equation}

\noindent and we impose a floor of $\mathrm{log} \, \Upsilon_{*,\mathrm{F125W}} = -1$.

However, no F336W or F125W observations for DYNAMO C13-1 were available; thus, we use HST ACS/WFC FR647M instead and derive the masses from:

\begin{equation}
    \mathrm{log} \, M_{*} = \mathrm{log}(F_\mathrm{FR647M} \; [\mathrm{Jy\,cm^{2}}]) - 42.04.
\end{equation}

\subsection{Molecular Gas Surface Density} \label{subsec:gsd}
We measure the molecular gas surface density from our integrated intensity maps of CO(3$-$2) in all cases but DYNAMO D15-3, for which no CO(3$-$2) was available. In that case, we use the CO(4$-$3) integrated intensity map. For each beam-sized aperture in our ``grid'', we extract the median integrated intensity of all pixels within the aperture and calculate the molecular gas surface density from: 

\begin{equation}
    \Sigma_{\mathrm{mol}} = \alpha_{\mathrm{CO}} \times R_{J,J-1} \times I_{CO} \;\; [\mathrm{M_{\odot}\,pc^{-2}}]
\end{equation}

\noindent where $R_{J,J-1}$ is the conversion of the CO(J$\rightarrow$J$-$1) emission to \jupone{}, for which we adopt the $R_{31}$ and $R_{41}$ values in Table 3 of \citet{lenkic23}, and $I_{CO}$ is the CO(J$\rightarrow$J$-$1) integrated intensity in units of K\,km\,s$^{-1}$. 

In the constant CO-to-H$_2$ conversion factor case, we adopt \aco $= 4.35$~M$_{\odot}$\,[K\,km\,s$^{-1}$\,pc$^{2}$]$^{-1}$, which represents the average value for a Milky Way-like galaxy \citep[e.g.,][]{bolatto13}. In the variable \aco{} case, we adopt the prescription of \citet{bolatto13} in their equation 31:

\begin{equation}
    \alpha_{\mathrm{CO}} \sim 2.9 \times \mathrm{exp}\left(\frac{0.4}{Z'\,\Sigma_{\mathrm{GMC}}^{100}}\right) \times \left(\frac{\Sigma_{\mathrm{total}}}{100\,\mathrm{M_{\odot}\,pc^{-2}}}\right)^{\gamma}
    \label{eq:var_aco}
\end{equation}

\noindent where $Z'$ is the metallicity normalized to the solar value, $\Sigma_{\mathrm{GMC}}^{100}$ is the gas surface density in units of 100~M$_{\odot}$\,pc$^{-2}$, $\Sigma_{\mathrm{total}}$ is the gas plus stellar surface density, and $\gamma = 0.5$ for $\Sigma_{\mathrm{total}} > 100$~M$_{\odot}$\,pc$^{-2}$ and 0 otherwise. We use an iterative approach to determine \aco{} for each beam-sized line-of-sight region by (1) calculating the initial gas surface density with \aco $= 4.35$~M$_{\odot}$\,[K\,km\,s$^{-1}$\,pc$^{2}$]$^{-1}$, (2) deriving \aco{} based on equation \ref{eq:var_aco}, (3) recalculating the gas surface density with the updated \aco{}, and (4) repeating this process until \aco{} changes by less than 0.1\%.

\subsection{Star Formation Rate Surface Density}
To obtain SFR surface densities, we use the HST H$\alpha$ observations matched to the resolution and pixel scale of the CO data. We measure the H$\alpha$ flux along each beam-sized line-of-sight aperture in our two grids. We convert these fluxes to units of erg\,s$^{-1}$\,cm$^{-2}$\,\AA$^{-1}$ and apply a correction for extinction \citep[see][for details]{lenkic23}. Finally, we calculate H$\alpha$ luminosities and convert them to SFRs using the relation of \citet{hao11}:

\begin{equation}
    \mathrm{SFR\;[M_{\odot}\,yr^{-1}]} = 5.53 \times 10^{-42} \times L_{\mathrm{H}\alpha} \; [\mathrm{erg\,s^{-1}}].
\end{equation}

The global \sfrsd{} values shown in Table \ref{tab:properties} are from \citet{fisher19}, and are derived from their global SFR and $R_{1/2}$ measurements (see their Tables 1 and 2).

\subsection{Velocity Dispersion} \label{subsec:gas_disp}
To test theories of star formation, we finally must derive molecular gas velocity dispersions, which are considered to trace turbulence in the ISM. The ALMA CO observations allow us to measure the molecular gas velocity dispersion on $1-2$~kpc scales and to compare these as a function of molecular gas and SFR surface densities to model predictions, which we will show in \S\ref{subsec:vdsip_molsd}. Here, we outline our method for measuring molecular gas velocity dispersions and correcting for beam smearing.

To measure the velocity dispersion, we use the \jupthree{} observations when available, and the \jupfour{} observations otherwise. We begin by creating two overlapping grids of beam-sized apertures as described in \S\ref{sec:methods}. For each aperture, we then extract the spectrum of the CO line from the central pixel. We fit the line profile with a Gaussian function of the form $f(x) = a \times e^{(x-\mu_{o})^{2}/2\sigma^{2}}$, where $a$ is the amplitude of the line, $\mu_{o}$ is its centroid, and $\sigma$ is its velocity dispersion. We perform our fitting using the {\sc Python SciPy} function \texttt{curve\_fit}. The resulting $\sigma$ parameters obtained in this way are our velocity dispersion measurements.

\begin{figure*}
    \centering
    \includegraphics[width=\textwidth]{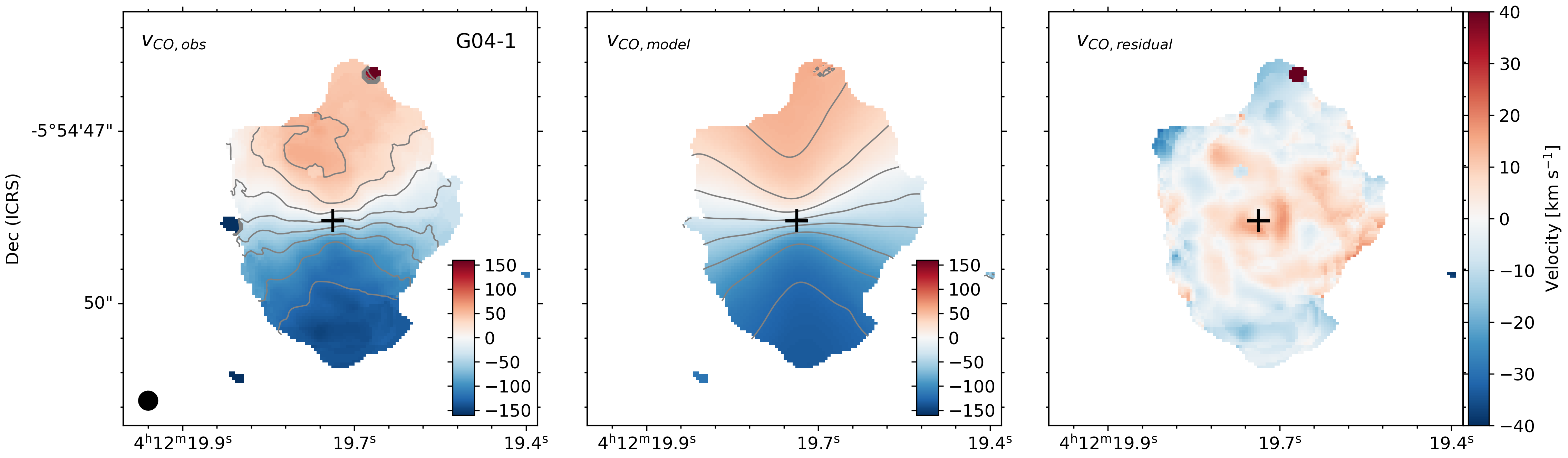}
    
    \includegraphics[width=\textwidth]{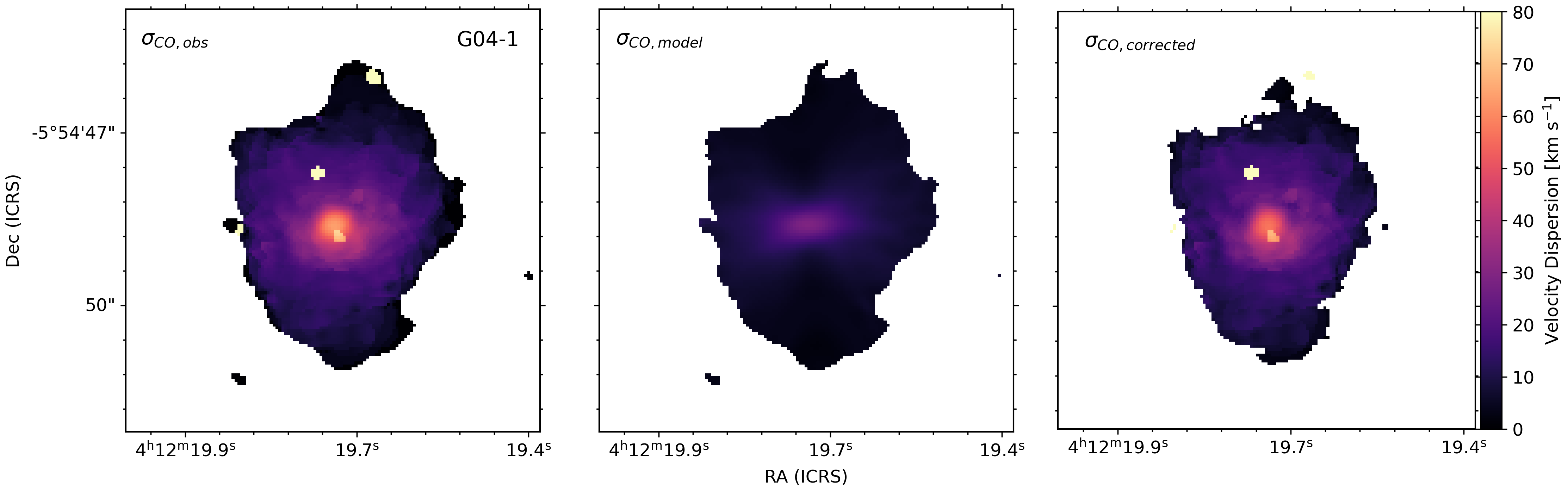}
    \caption{Observed and modeled velocity field and velocity dispersion maps of DYNAMO G04-1. \textit{Left:} the observed velocity field (top) and velocity dispersion (bottom). \textit{Middle:} the modeled velocity field (top) and velocity dispersion (bottom), derived from a model data cube. \textit{Right:} The velocity field residuals we obtain by subtracting the model velocity field from the observations (top), and the corrected velocity dispersion map obtained by subtracting in quadrature the model velocity dispersion from our measured velocity dispersion. This simulates the effect of beam smearing and corrects for it.}
    \label{fig:beam_smear}
\end{figure*}

However, beam smearing has a strong effect on measured velocity dispersion, particularly close to the centers of galaxies and along their minor axes. The effect of beam smearing can be significant even into the disk of the galaxies if their rotation curves rise slowly \citep{Leung18}. As a result, we apply a beam smearing correction to our measured velocity dispersions using the method of subtracting in quadrature an estimated value of the velocity dispersion due to beam smearing alone. We follow the procedure outlined in \citet{levy18}, which we summarize here for completeness. To determine this correction, we first create model data cubes with no intrinsic dispersion. To accomplish this, we adopt the arctan model and rotation curve parameters from \citet{girard21} to calculate the rotation velocity ($v_{rot}$) as a function of distance from the galaxy center ($r$):

\begin{equation}
    v_{rot}(r) = \frac{2}{\pi} \times V_{rot} \times \mathrm{arctan}\left(\frac{r}{r_{t}}\right)
\end{equation}

\noindent where $V_{rot}$ is the rotation velocity in the flat region of the rotation curve and $r_{t}$ is the ``turnover radius'' where the rotation curve transitions from rising to flat. We then calculate the observed velocity (i.e., in the plane of the sky; $v_{obs}$) from:

\begin{equation} \label{eq:v_obs}
    v_{obs}(x,y) = v_{rot}(r) \times \mathrm{sin}(i) \times \mathrm{cos}(\theta)
\end{equation}

\noindent where $i$ is the inclination \citep[from][]{girard21} and cos($\theta$) is defined as:

\begin{equation}
    \mathrm{cos}(\theta) = \frac{-(x - x_{o}) \times \mathrm{sin}(\phi) - (y - y_{o}) \times \mathrm{cos}(\phi)}{r \times \mathrm{cos}(i)}
\end{equation}

\noindent where $x_{o}$ and $y_{o}$ are the coordinates of the galaxy center and $\phi$ is the galaxy position angle \citep[see e.g.,][]{Begeman1989}. Finally, $r$ is defined in the plane of the galaxy as:

\begin{equation}
    r = \sqrt{\frac{(x - x_{o})^{2} + (y - y_{o})^{2}}{\mathrm{cos}(i)^{2}}}.
\end{equation}

We build our model cubes by creating an array of zeros with the same dimensions as the actual observed cube, and then calculating $v_{obs}$ at each pixel using equation (\ref{eq:v_obs}). We match the resulting observed velocity to the closest channel and place a ``line'' (a delta function with a linewidth equal to one channel) in the model cube array at the corresponding pixel and channel. The amplitude of the line is equal to the brightness of the same pixel in the same channel of the observed data cube (in Jy\,beam$^{-1}$). After performing this for each pixel, we smooth the cube with a Gaussian whose FWHM matches the resolution of the observation. Then, we fit the model CO line profile with a Gaussian, as above, to estimate the velocity dispersion due to beam smearing. Finally, we apply the beam smearing correction by subtracting the model velocity dispersion  ($\sigma_{\mathrm{CO,model}}$) from the observed velocity dispersion ($\sigma_{\mathrm{CO,observed}}$) in quadrature for each beam-sized line-of-sight aperture to obtain final beam smearing corrected velocity dispersion ($\sigma_{\mathrm{CO,corrected}}$) measurements.

In Figure \ref{fig:beam_smear}, we show the steps of this correction for DYNAMO G04-1. In the top panel of Figure \ref{fig:beam_smear}, we show the observed (left panel) and modeled (middle panel) moment 1 maps of DYNAMO G14-1, and the residuals ($v_{obs} - v_{mod}$) in the rightmost panel. The residuals for DYNAMO G04-1 show a pattern of positive residuals that resemble the spiral arms of this galaxy. The bottom panels of Figure \ref{fig:beam_smear} show the observed (left panel) and modeled (middle panel) velocity dispersions, while the rightmost panel shows the observed velocity dispersion map after beam smearing correction ($\sqrt{\sigma_{obs}^{2} - \sigma_{mod}^{2}}$).

The global \smol{} values we report in Table \ref{tab:properties} are from \citet{girard21}. The authors model the ALMA CO data cubes with GalPak$^{3D}$ \citep{bouche15}. The cubes are fit directly with an arctan function and velocity dispersion is assumed to be constant across the disk. The model is then convolved with the beam and line spread function, which accounts for beam smearing.

\section{Results} \label{sec:results}
Our sample consists of nearly ${\sim} 500$ kiloparsec-scale measurements of the star forming, stellar and gas mass, and velocity dispersion properties across seven DYNAMO galaxies. Because DYNAMO galaxies resemble $z \sim 1$ star forming systems, we can now investigate the relationships between these quantities and what they reveal about star formation regulation at physical scales not yet achievable in the high-redshift Universe. 

\subsection{Galaxy Averaged Relationship between \smol{}, $\sigma_{\mathrm{ion}}$, and \sfrsd{}}
\begin{figure*}
\begin{center}
\includegraphics[width=\textwidth]{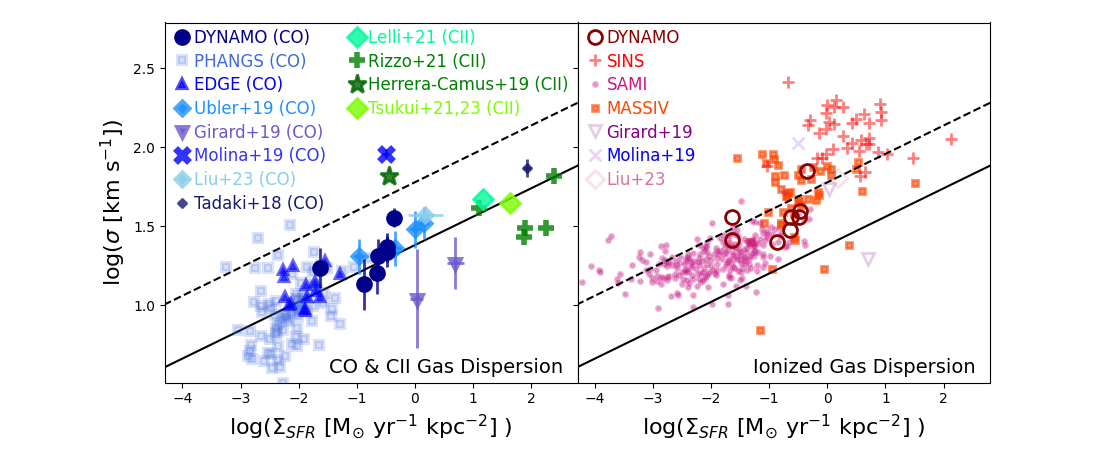}
\end{center}
\caption{Global values of velocity dispersion, using the median dispersion from the moment maps, and \sfrsd{} are compared to several samples for both cold gas tracers (CO and \CII{}) on the {\bf left} and ionized gas on the {\bf right}. There is a significant correlation between \sfrsd{} and velocity dispersion, with a systematic offset to higher dispersion when measured with ions. In both panels, the solid line indicates a fit to \sfrsd{} and \smol{}, and the dashed line represents $2.5\times\sigma_{mol}$. CO data from literature sources are taken from PHANGS \citep[$z\sim0$;][]{Sun2023}, EDGE \citep[$z = 0.005-0.03$;][]{levy18}, \cite{Ubler2019} ($z\sim2$), \cite{Girard2019} ($z\sim1-1.5$), \cite{Molina2019} ($z\sim1.5$), \cite{Liu2023} ($z\sim2$), \cite{Tadaki2018} ($z\sim4$); for \CII{} they are taken from \citet{Lelli2021} ($z \sim 5$), \citet{Rizzo2021} ($z \sim 4-5$), \citet{HerreraCamus2022} ($z \sim 5.5$), \citet{Tsukui2021,Tsukui2023} ($z \sim 4.4$). Ionized gas velocity dispersion measurements are taken from SAMI \citep[$z\sim0$;][]{Varidel2020}, MASSIV \citep[$z\sim1-1.5$;][]{Epinat2012}, and SINS \citep[$z\sim2$;][]{forsterschreiber11}.}  \label{fig:globalsig}
\end{figure*}

In Figure~\ref{fig:globalsig}, we compare the velocity dispersion and \sfrsd{} of DYNAMO galaxies to several other samples, spanning a range of systems from local spirals to $z~\sim~5$ galaxies. We show comparisons to both ionized (right panel) and molecular gas (left panel).

Several studies have discussed the similarities of ionised gas velocity dispersions of DYNAMO galaxies to $z\sim1$ galaxies \citep{green14,Bassett2014,oliva-altamirano18}. In Figure~\ref{fig:globalsig}, we reiterate this result. \cite{fisher19} compiled a list of high-quality ionised gas velocity dispersion measurements for DYNAMO galaxies using data from Gemini/GMOS and Keck/OSIRIS. The average $\sigma_{\mathrm{ion}}$ from that sample of 17 galaxies is 47~\kms{}. The galaxy averaged velocity dispersions and \sfrsd{} of DYNAMO galaxies are most similar to galaxies from the MASSIV sample \citep[$z \sim 1$;][]{Epinat2012}, where the average $\sigma_{\mathrm{ion}}$ is 53~\kms{}. In comparison to $z \sim 0$ galaxies from the SAMI survey \citep{Varidel2020}, the DYNAMO velocity dispersions are over twice as high as the average of the SAMI sample, where $\sigma_{\mathrm{ion}} \sim 20$~\kms{}.

There are significantly fewer measurements of molecular gas velocity dispersion for comparison; nevertheless, for the data that exist in the literature we find a similar result. \cite{girard21} finds that the average \smol{} for DYNAMO galaxies is $\sim$13~\kms{}. In contrast, the average \smol{} for galaxies in the PHANGS sample is $\sim$7~\kms{}. Because the inclination can increase the measured velocity dispersion, this average corresponds to PHANGS galaxies with inclinations ${<}50^{\circ}$, which is similar to our DYNAMO sample. For comparison to $z\sim1-2$ galaxies, we combine the samples of \cite{Ubler2019} and \cite{Girard2019}, noting the caveat that this results results in a sample of only six heterogeneously selected galaxies. The typical \smol{} for this sample of six $z \sim 1$ targets is ${\sim} 20$~\kms{}. 

In Figure~\ref{fig:globalsig}, we also show a fitted curve between \smol{} and \sfrsd{} for the full sample of galaxies in the literature (we have weighted the PHANGS galaxies down by a factor of five, so that they do not dominate the minimization). The sources of comparison include: CO observations of $z=0$ galaxies from the PHANGS \citep{Sun2023} and EDGE \citep{levy18} samples; CO observations of $z\sim1-2$ galaxies from PHIBBS \citep{Ubler2019}, 3 lensed galaxies \citep{Girard2019,Liu2023}, and a single target from the SHiZELS sample \citep{Molina2019}. At higher redshift, the only target with CO velocity dispersion measurements is AzTEC-1 \citep{Tadaki2018}; therefore, we also add galaxies in which the velocity dispersion is traced by \CII{} \citep{Rizzo2021, Lelli2021,HerreraCamus2022}. We offer the strong caveat that the sample is not homogeneously selected and there is a difference in data quality from the low$-z$ targets of DYNAMO, PHANGS, and EDGE to the high$-z$ targets. Nevertheless, without large ALMA programs, this is the only means to derive such correlations.

We find a correlation between the galaxy averaged molecular gas velocity dispersion and \sfrsd{}, such that 

\begin{equation}
    \log \sigma_{\mathrm{mol}} = (0.19\pm0.03) \, \times \, \log \Sigma_{\mathrm{SFR}} + (1.33\pm0.04).
    \label{eq:global_smol}
\end{equation}
In both panels of Figure~\ref{fig:globalsig}, we also show a dashed line that represents $2.5\times \sigma_{\mathrm{mol}}$. This is the scale-factor that was found in \cite{girard21} from the fitted relationship between \smol{} and $\sigma_{\mathrm{ion}}$, which appears to show overall agreement with the data here. We note that there are two very high \smol{} outliers to this power-law: They are SHiZELS-19 \citep{Molina2019} and HZ4 \citep{HerreraCamus2022}. Given the small number of targets at high \sfrsd{}, we are careful not to merely dismiss these as they may represent an important subset of galaxies at high$-z$, or point to differences in analysis techniques. More data is clearly needed to study this important power-law for galaxy properties.

The power-law that we find in Figure~\ref{fig:globalsig} is very similar to the power-law found in the recent SILCC simulations \citep{Rathjen2023}. The authors run a set of simulations with varying \gsd{} that incorporate feedback to drive the velocity dispersions. A similar fit between \smol{} and \sfrsd{} yields a slope of $0.2$. They find a constant offset between the warm (ionized) and cold (molecular) gas velocity dispersion of ${\sim} 2.2$. We note that the SILCC simulations do not incorporate large scale instabilities that are often invoked to explain large velocity dispersion, but rather drive the velocity dispersion only through a complex model of star formation feedback. They can in this model recover both the velocity dispersions of the ions and the molecular gas.

\subsection{\gsd{}$-$\ssd{} Relation} \label{subsec:ssd_gsd}
\begin{figure}
    \centering
    \includegraphics[width=\columnwidth]{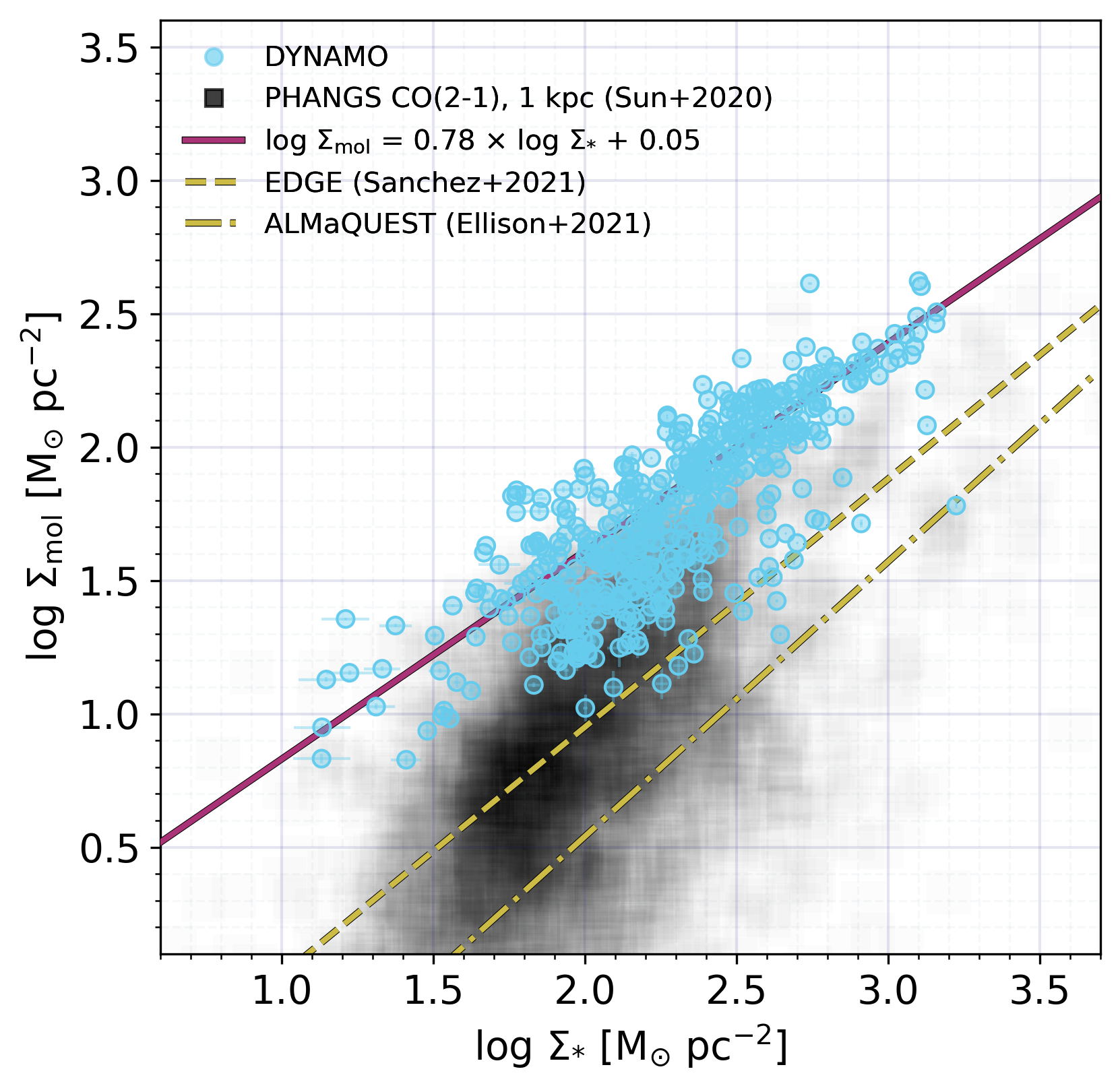}
    \caption{The molecular gas mass surface density as a function of stellar mass surface density. The blue data points correspond to DYNAMO ${\sim} 1-2$~kpc line-of-sight measurements, assuming a variable \aco{} conversion factor as described by equation \ref{eq:var_aco}. For comparison, we include the PHANGS \juptwo{} ${\sim} 1$~kpc line-of-sight measurements as black data points, and best fit lines from local galaxy samples in EDGE-CALIFA \citep[yellow dashed line;][]{barrera-ballesteros20,sanchez21} and ALMaQUEST \citep[yellow dash-dotted line;][]{Ellison2021aq}. We also perform an ODR fit to the DYNAMO measurements (solid purple line), which shows that DYNAMO galaxies are fitted by a shallower slope and larger normalization. DYNAMO galaxies are fundamentally more gas-rich than local spirals at all disk positions.}
    \label{fig:ssd_gsd}
\end{figure}

In Figure~\ref{fig:ssd_gsd}, we show the relationship between \gsd{} and \ssd{}. DYNAMO measurements are presented as blue circles, while the black squares correspond to measurements from the PHANGS-ALMA survey \citep{sun20,leroy21}, which have been smoothed to kpc-scale resolution to match our observations. PHANGS-ALMA observed 90 nearby (d\,$\lesssim$\,20~Mpc) galaxies that are on or near the $z = 0$ main-sequence in \juptwo\ at ${\sim} 100$~pc resolution, and \citet{sun20} present results for 70 of these targets, consisting of 102,778 independent lines of sight. The data we show in Figure \ref{fig:ssd_gsd} and subsequent ones include only lines of sight from galaxies with inclinations less than 50$^{\circ}$ (corresponding to the largest inclination in our sample). The \gsd{}$-$\ssd{} relationship is sometimes referred to as the ``molecular gas main-sequence''. It has been measured for several samples in nearby galaxies \citep{barrera-ballesteros20,ellison21,sanchez21,lin19}, and is often thought to drive the star formation main-sequence, in combination with the KS relation \citep{Ellison2021aq,Baker2023}, although other studies find no conclusive evidence for this \citep[see e.g.,][]{sanchez21}. \cite{Ellison2021aq} showed that of the correlations between \sfrsd{}, \ssd{}, and \gsd{}, the tightest relationship is between \ssd{} and \gsd{}, with a scatter of ${\sim} 0.19$~dex in their ALMaQUEST sample. Moreover, the scatter in the \ssd{}$-$\gsd{} relationship does not correlate to scatter in the KS relation. We can therefore use this as an independent means of comparing resolved \gsd{} in DYNAMO galaxies to that of local spirals. We note there are not a sufficient number of galaxies observed at $z>1$ for a similar comparison.

We show in Figure~\ref{fig:ssd_gsd} that an orthogonal distance regression (ODR) fit to the resolved regions in DYNAMO galaxies yields a sub-linear relationship that is offset to higher \gsd{} than the relationships derived on local spirals \citep{barrera-ballesteros20,ellison21,sanchez21,lin19}. The relationship we determine for DYNAMO galaxies is: 

\begin{equation}
    \log \Sigma_{\mathrm{mol}} = (0.78\pm0.03) \times  \log \Sigma_{*} + (0.05\pm0.07).
\end{equation}
The extrapolation of this relationship toward low \ssd{} does not project into the sequence of PHANGS galaxies. This brings up a useful insight into the nature of DYNAMO galaxies. They are not a continuation of the properties of local Universe spirals, nor are they in this way similar to the centers of local spirals. They are globally more gas rich at all positions in the disk. DYNAMO galaxies have \gsd{}/\ssd{} that is roughly an order-of-magnitude higher than local spirals at all values of \ssd{} observed. This is consistent with previous global measurements of DYNAMO galaxy gas fractions being higher than local Universe galaxies \citep{fisher14,White2017,fisher19}. 

\subsection{Resolved Molecular KS Relation at High \sfrsd{}} \label{subsec:ks_law}
\begin{figure*}
    \centering
    \includegraphics[width=\textwidth]{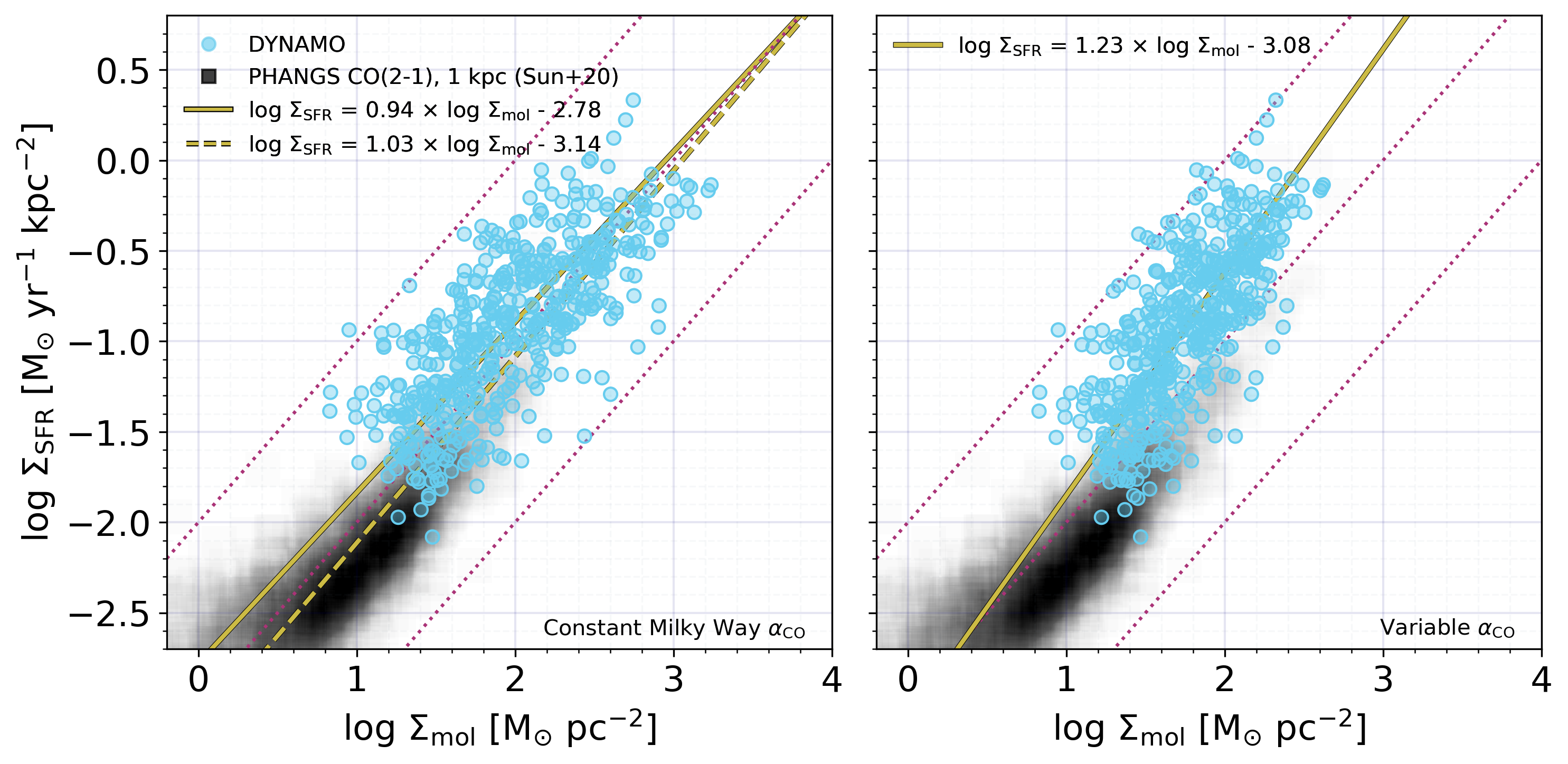}
    \caption{The kpc-scale resolved Kennicutt-Schmidt relation for DYNAMO galaxies with \gsd{} measured assuming a constant \aco{} (left panel) and a varying \aco{} (right panel). We include the $\sim$1~kpc scale measurements of \citet{sun20} for PHANGS galaxies, using \juptwo{} maps and H$\alpha$+24~$\mu$m derived SFRs, as black squares. The solid yellow line shown in both panels is the best-fit line (determined using orthogonal linear regression) to the DYNAMO measurements. The dashed yellow line is the best-fit relation to the DYNAMO and PHANGS data combined. The dotted purple lines indicate constant molecular gas depletion times of 0.1, 1, and 10~Gyr (from top left to bottom right).}
    \label{fig:ks_law}
\end{figure*}

We have shown that the DYNAMO galaxies in our sample are systematically more gas rich than local star-forming galaxies; this allows us to derive the KS relation at kiloparsec scales for systems that are selected to resemble $z \sim 1$ galaxies, where such resolved observations are still challenging. This is what we present in Figure \ref{fig:ks_law}: The left panel assumes a constant Milky Way \aco{} for deriving \gsd{}, while the right panel assumes a variable \aco{} (see \S\ref{subsec:gsd}). Measurements of \gsd{} and \sfrsd{} by \citet{sun20} from the PHANGS-ALMA \juptwo{} survey \citep{leroy21}, matched to the kpc-scale resolution of our observations, are included as black squares.

We fit our measurements assuming both a constant and variable \aco{} with a power law of the form $\mathrm{log}$\,\sfrsd\,$= N\, \times\, \mathrm{log}$\,\gsd\,$+\,C$ using ODR. In the constant \aco{} case, we find 

\begin{equation}
    \mathrm{log}\;\Sigma_{\mathrm{SFR}} = (0.90 \pm 0.04) \, \times \, \mathrm{log} \; \Sigma_{\mathrm{mol}} - (2.70 \pm 0.08)
\end{equation}

\noindent when fitting the DYNAMO measurements alone \citep[which is very similar to the slope of $N = 0.92$ from][when assuming a MW \aco{}]{Sun2023}, and 

\begin{equation}
    \mathrm{log}\;\Sigma_{\mathrm{SFR}} = (1.04 \pm 0.01) \, \times \, \mathrm{log} \; \Sigma_{\mathrm{mol}} - (3.15 \pm 0.02)
\end{equation}

\noindent when fitting the DYNAMO and PHANGS measurements together. In the variable \aco{} case, we find

\begin{equation}
    \mathrm{log}\;\Sigma_{\mathrm{SFR}} = (1.23 \pm 0.03) \, \times \, \mathrm{log} \; \Sigma_{\mathrm{mol}} - (3.08 \pm 0.06)
\end{equation}

\noindent for the DYNAMO data, which is very similar to the slope of $N = 1.21$ from \citet{Sun2023} when they assume the same variable \aco{} prescription.

The near-unity slopes we find in the constant Milky Way \aco{} case would suggest that the depletion time (\tdep{}\,$=$\,\gsd{}/\sfrsd{}) in DYNAMO is roughly constant. The purple dotted lines in Figure~\ref{fig:ks_law} indicate constant depletion times of 0.1, 1, and 10~Gyr from top left to bottom right. We can see from this that more than half of the DYNAMO line-of-sight measurements (297/490) have \tdep{}\,$<1$~Gyr. In fact, the median depletion time and 16th$-$84th percentile ranges we find are \tdep{}\, $=0.8${\raisebox{0.5ex}{\tiny$^{+0.9}_{-0.4}$}}~Gyr. This is significantly lower than \citet{Sun2023} who find a median \tdep{}\,$= 2.1${\raisebox{0.5ex}{\tiny$^{+1.9}_{-1.1}$}}~Gyr when they assume a constant Milky Way \aco{} in over 2000 kpc-sized apertures across the full sample of 80 galaxies from PHANGS-ALMA.

When we adopt an \aco{} which varies with local gas plus stellar mass surface density \citep{bolatto13}, we find a steeper slope of $N = 1.23$, suggesting that depletion time is not constant in DYNAMO, and becomes shorter at higher \gsd{} and \sfrsd{}. In this case, nearly all DYNAMO regions have depletion times shorter than 1~Gyr (415/490), and the median depletion time is \tdep{}\,$= 0.5${\raisebox{0.5ex}{\tiny$^{+0.5}_{-0.3}$}}. Similarly, in a case study of the nearby ($z \sim 0.02$) high-redshift galaxy analog IRAS08339+6517, \citet{Fisher2022} find two orders-of-magnitude variation in depletion time with \tdep{}\,$< 0.1$~Gyr in the central kiloparsec and \tdep{}\,$> 3$~Gyr at radii greater than ${\sim} 2.5$~kpc. Conversely, \citet{Sun2023} find a median \tdep{}\,$= 1.9${\raisebox{0.5ex}{\tiny$^{+1.5}_{-1.0}$}} when assuming the variable \aco{} prescription of \citet{bolatto13}.

\begin{deluxetable*}{lccc}
\tablecaption{Summary of ODR Power Law Fit Results}
\label{tab:obs}
\tablewidth{700pt}
\tabletypesize{\normalsize}
\tablehead{
	\colhead{Relation}                         &
	\colhead{Power Law Index, $N$}              &
	\colhead{Intercept, $C$}                    \\
} 
\startdata
Galaxy averaged \smol{}$-$\sfrsd{} & 0.19 $\pm$ 0.03 & 1.33 $\pm$ 0.04 \\
Resolved \gsd{}$-$\ssd{} & 0.78 $\pm$ 0.04 & 0.05 $\pm$ 0.07 \\
Resolved KS Law, constant \aco{}, DYNAMO only & 0.90 $\pm$ 0.04 & $-$2.70 $\pm$ 0.08 \\
Resolved KS Law, constant \aco{}, DYNAMO+PHANGS & 1.04 $\pm$ 0.01 & $-$3.15 $\pm$ 0.02 \\
Resolved KS Law, variable \aco{}, DYNAMO only & 1.23 $\pm 0.03$ & $-$3.08 $\pm$ 0.06 \\
Resolved \smol{}$-$\gsd{}, DYNAMO+PHANGS & 0.48 $\pm$ 0.02 & 0.47 $\pm$ 0.03 \\
Resolved \smol{}$-$\sfrsd{}, DYNAMO+PHANGS & 0.27 $\pm$ 0.02 & 1.56 $\pm$ 0.02 \\
Resolved \smol{}$-$\tdep{} & -0.80 $\pm$ 0.07 & 1.00 $\pm$ 0.03
\enddata
\end{deluxetable*}

In addition to PHANGS-ALMA, there are several additional surveys of local  galaxies that have measured the molecular KS relation on kpc scales which we can compare to. \citet{Ellison2021aq} use 15,000 kpc-sized spaxels across 28 galaxies ($0.02 < z < 0.05$) from ALMaQUEST to find a slope of $N = 1.23 \pm 0.01$ when assuming a Milky Way \aco{}. This is higher than our slope of $N = 0.94$ under the same assumption for \aco{}. For comparison, the authors also assume the metallicity-dependent \aco{} prescription of \citet{sun20} and find a slope of $N = 1.27$. \citet{sun20} shows that the variable \aco{} prescription of \citet{bolatto13} leads to a higher slope than their fiducial metallicity-dependent \aco{}. In contrast, \citet{sanchez21} find a slope of $0.98 \pm 0.14$ for ${\sim} 15,500$ kpc-scale line-of-sight measurements from EDGE-CALIFA for a constant Milky Way CO-to-\Htwo{} conversion factor. For ${\sim} 14,500$ kpc-scale measurements across a sample of 30 nearby disk galaxies from HERACLES \citep{leroy09}, \citet{leroy13} find a slope of $N = 1.00 \pm 0.15$ for the KS relation when assuming a Milky Way \aco{}, consistent with our results. However, they find that the median gas depletion time is 2.2~Gyr with 0.3~dex scatter, consistent with the PHANGS-ALMA results, but longer than our depletion time of \tdep{}\, $=0.8${\raisebox{0.5ex}{\tiny$^{+0.9}_{-0.4}$}}~Gyr when assuming a constant Milky Way \aco{}.

\subsection{Resolved Correlations of \smol{}$-$\gsd{}, \smol{}$-$\sfrsd{}, and \smol{}$-$\tdep{}} \label{subsec:vdsip_molsd}
\begin{figure*}
    \centering
    \includegraphics[width=\textwidth]{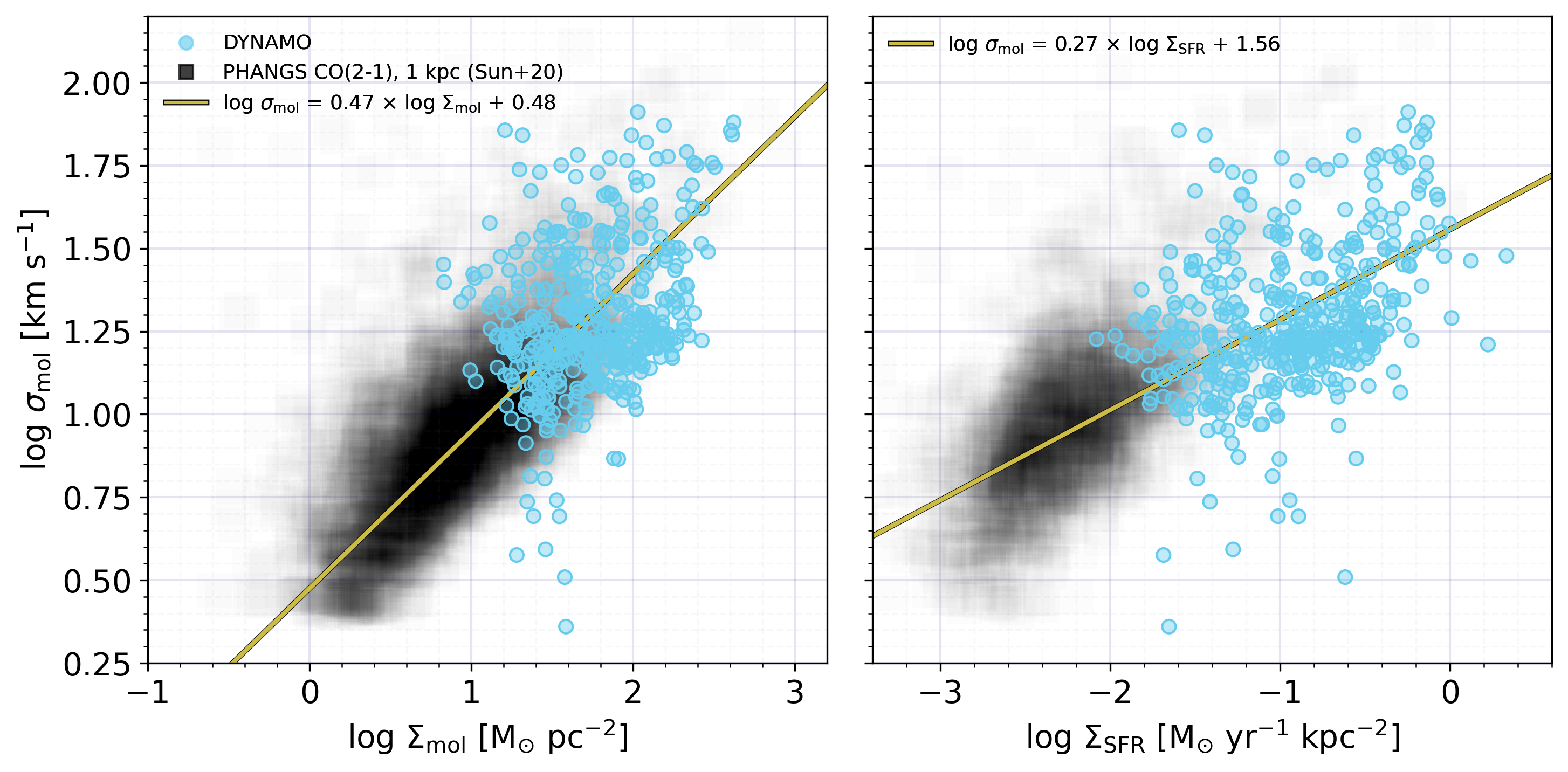}
    \caption{Beam smearing corrected molecular gas velocity dispersions as a function of molecular gas surface density (left), and star formation rate surface density (right), for individual $\sim 1-2$~kpc-scale lines of sight measurements in seven DYNAMO galaxies. The black squares correspond to the PHANGS-ALMA derived measurements of \citet{sun20} in 70 nearby galaxies, at a resolution of 1~kpc to match our observations (J. Sun, private communication). Consistent with the PHANGS-ALMA results, velocity dispersions in DYNAMO galaxies increase with \gsd{} and \sfrsd{}. In contrast to PHANGS, DYNAMO galaxies populate the high \smol{}, \gsd{}, and \sfrsd{} regime of this parameter space. Furthermore, these high values are observed throughout the entire disks of DYNAMO galaxies, not just their central regions.}
    \label{fig:disp_sh2}
\end{figure*}

Figure \ref{fig:disp_sh2} presents the \smol{}$-$\gsd{} relation in the left panel, and the \smol{}$-$\sfrsd{} relation in the right panel, where all measurements are made along ${\sim} 1-2$~kpc-sized apertures. As in previous figures, DYNAMO data points are in blue, while the kpc-scale matched resolution measurements of \smol{}, \gsd{} and \sfrsd{} by \citet{sun20} from PHANGS-ALMA are included as black squares.
In both panels, the solid yellow line is an ODR power-law fit to both the DYNAMO and PHANGS measurements:

\begin{equation}
    \mathrm{log}\;\sigma_{\mathrm{mol}} = (0.48 \pm 0.02) \, \times \, \mathrm{log\;}\Sigma_{\mathrm{mol}} + (0.47 \pm 0.03)
\end{equation}

\noindent and

\begin{equation}
    \mathrm{log}\;\sigma_{\mathrm{mol}} = (0.27 \pm 0.2) \, \times \, \mathrm{log\;}\Sigma_{\mathrm{SFR}} + (1.56 \pm 0.02).
\end{equation}

We make several observations from Figure \ref{fig:disp_sh2}. First, we note that the results of our \smol$-$\gsd\ relation are consistent with the kpc-scale measurements of \citet{sun20}: regions of more active star formation tend to host molecular gas with higher surface densities. \citet{sun20} compare the centers of barred galaxies to disk regions and find that barred centers have ${\sim} 20\times$ higher mass-weighted molecular gas surface densities and $5\times$ higher mass-weighted median molecular gas velocity dispersions. The authors attribute these high \gsd{} and \smol{} values to the presence of stellar bars, which drive large-scale gas inflows, boosting \gsd{}, and enhance local turbulence through the release of gravitational potential energy. 

We also see in the left panel of Figure \ref{fig:disp_sh2} that despite being local galaxies, DYNAMO systems are unlike the PHANGS-ALMA nearby targets, because DYNAMO galaxies have high \smol{} and \gsd{} everywhere in their disks (though the centers of DYNAMO do also exhibit higher \smol{}; see Figure \ref{fig:rad_disp}). The DYNAMO galaxies we study are not barred in near-IR starlight images. However, they have been shown to have low values of Toomre$-Q$, indicating galaxy wide instabilities \citep{fisher17b,White2017}. Such instabilities are likewise associated to inflows in disks \citep{Dekel2014}, which is seen in the nearby clumpy, blue compact disk galaxy IRAS08339+6517 \citep{Fisher2022}.

\begin{figure}
    \centering
    \includegraphics[width=\columnwidth]{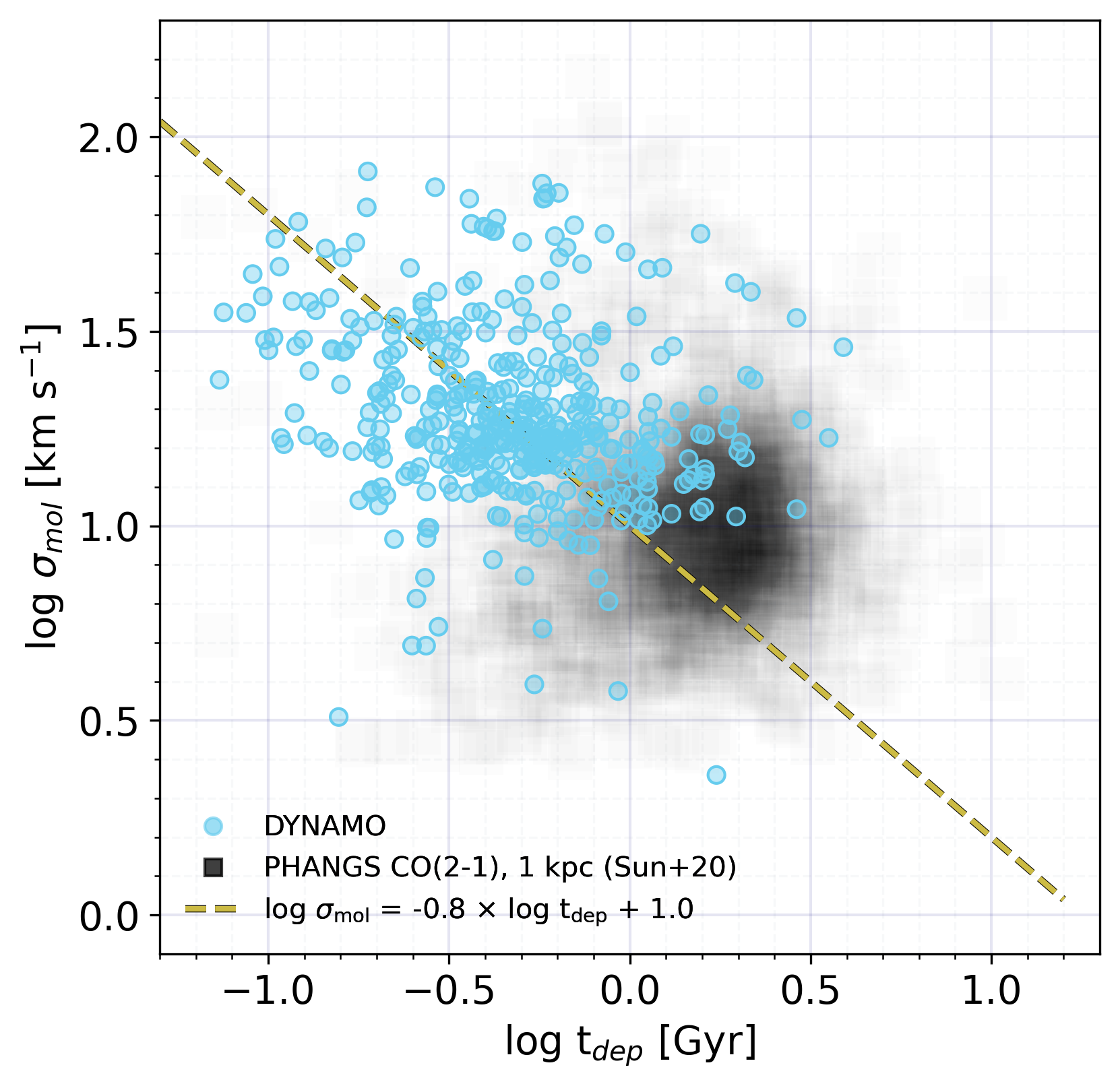}
    \caption{Beam smearing corrected molecular gas velocity dispersion as a function of the gas depletion time (\tdep{}), measured along individual ${\sim} 1-2$~kpc sized lines of sight and assuming a variable \aco{}. Compared to PHANGS, DYNAMO disks have overall shorter depletion times. As we would expect, the shorter depletion times are at smaller galactocentric radii, where the velocity dispersions are greater. The yellow dashed line represents the best fit to the DYNAMO+PHANGS measurements.}
    \label{fig:tdep}
\end{figure}

In Figure \ref{fig:tdep}, we plot the molecular gas velocity dispersion as a function of depletion time, assuming a variable \aco{}. We find an inverse relation between these two quantities, as was found by \citet{fisher19} for galaxy averaged values of \smol{} and \tdep{} for a sample of 14 galaxies, including 10 DYNAMO galaxies. The authors found that \smol{}\,$\propto$\,\tdep{}$^{-0.72}$, while the best-fit to our kpc-scale resolved DYNAMO measurements combined with PHANGS is:

\begin{equation}
    \mathrm{log}\;\sigma_{\mathrm{mol}} = (-0.80 \pm 0.07) \, \times \, \mathrm{log\;}t_{\mathrm{dep}} + (1.00 \pm 0.03).
\end{equation}

Feedback-regulated models of star formation \citep[see e.g.,][]{ostriker11,shetty12, faucher-giguere13}, where the gravitational force of the gas in the disk balances the momentum-flux injected into the ISM by supernovae, predict that velocity dispersion is linearly inversely proportional to the depletion time; i.e., \smol{}\,$\propto$\,\tdep{}$^{-1}$. 

There are, however, differences in different implementations of feedback-regulation. \citet{faucher-giguere13} argues that the important time-scale for turbulence to dissipate within the disk is the time relating to a circular orbit, where \cite{ostriker22} argues it is the vertical crossing time related to the disk thickness. We combine equations (6) and (18) in \citet{faucher-giguere13} to predict that:

\begin{equation}
    \sigma \propto (P_{*}/m_{*}) \times \frac{\Sigma_{\mathrm{SFR}}}{\Omega \, \Sigma_{\mathrm{gas}}} = (P_{*}/m_{*}) \times \frac{t_{orb}}{t_{dep}}.
    \label{eq:fg13}
\end{equation}

\noindent where $\Omega = v_{c}/r$ is the angular frequency, $t_{\mathrm{orb}} = \Omega^{-1}$ is the orbital time, and $P_{*}/m_{*}$ is the momentum returned to the ISM by stellar feedback per stellar mass formed. If we assume $P_{*}/m_{*} = 3,000$~\kms{} and adopt our measured values of \sfrsd{}, \gsd{}, and $\Omega$, then we are able to calculate the values of \smol{} that would be predicted by this simple feedback-regulated star formation model, and compare to our observed \smol{}$-$\sfrsd{} and \smol{}$-$\gsd{} relations. 

Taking the dynamical models of \cite{girard21}, it is straightforward to determine that $t_{orb}$ is shortest in the galaxy center, where we find both \gsd{} and \smol{} to be largest. This exercise reveals that the \smol{} values predicted by equation \ref{eq:fg13} decrease with increasing \gsd{} and \sfrsd{}, which is the opposite behavior we observe in Figure \ref{fig:disp_sh2}. The \smol{}\,$\propto$\,\tdep{}$^{-1}$ dependence in equation \ref{eq:fg13} is order-of-magnitude compatible with our \smol{}\,$\propto$\,\tdep{}$^{-0.8}$ dependence; therefore, we suggest that the assumption that turbulent momentum decay takes place on a eddy (disk) crossing time and is proportional to $\Omega$ is incompatible with our observations and is the cause of the discrepancy between the predicted \smol{} and our observed values.  

\section{Discussion} \label{sec:discussion}
\subsection{Comparison of \smol$-$\sfrsd\ and  \smol$-$\gsd\ to Hydrodynamic Models and Theory} \label{subsec:mod_comp}
\begin{figure*}
    \centering
    \includegraphics[width=\textwidth]{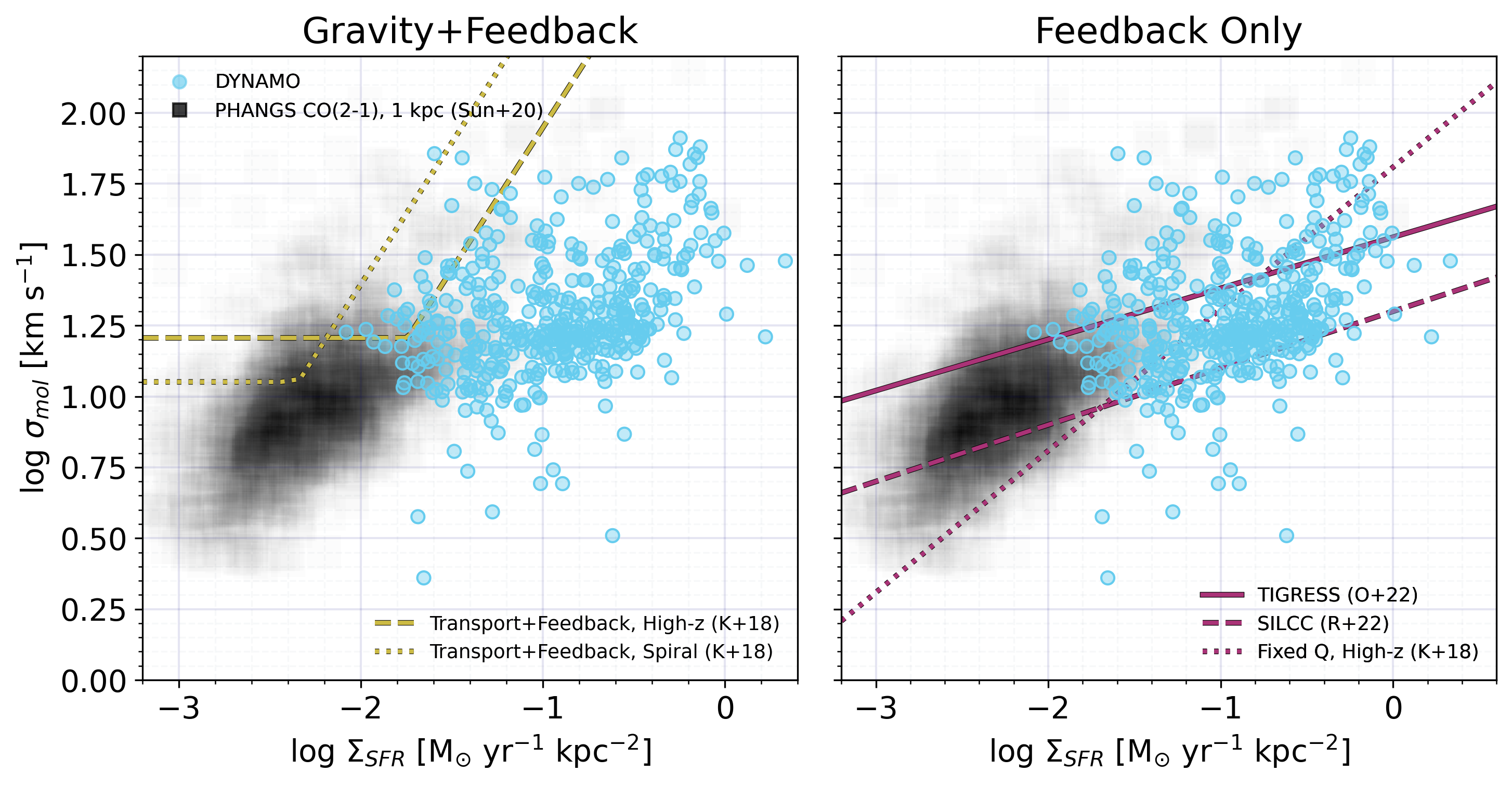}
    \caption{\textit{Left:} Comparison of DYNAMO (blue) and PHANGS (grey) \smol{} and \sfrsd{} measurements to the gas transport plus feedback model prediction of \citet{krumholz18}, where the yellow dashed line is the prediction for high-redshift galaxies and the yellow dotted line is for local spirals. This comparison shows that, for the fiducial parameters assumed in \citet{krumholz18}, the transport+feedback models overpredict the DYNAMO and PHANGS velocity dispersions. \textit{Right:} Comparison of DYNAMO and PHANGS measurements to the results of the TIGRESS \citep[solid purple line;][]{ostriker22} and SILCC \citep[dashed purple line;][]{Rathjen2023} simulations, which incorporate stellar feedback only, and the feedback-only model of \citet{krumholz18} assuming fixed $Q$ and variable $\epsilon_{\mathrm{ff}}$ (dotted purple line)}.
    \label{fig:sh2_comp}
\end{figure*}

In this section, we further explore our \smol{}$-$\sfrsd{} relation within the context of star formation regulation theories by comparing our observed relations to predictions from theory and results from numerical simulations. 

\subsubsection{Gravitational and Stellar Feedback-Driven Turbulence} \label{subsubsec:grav_fdbck}
In the left panel of Figure \ref{fig:sh2_comp}, we first compare the DYNAMO and PHANGS-ALMA results to the predictions of \citet{krumholz18}. They develop a model where the sources of turbulence in galaxy disks include both feedback from star formation and the release of gravitational potential energy from gas inflows \citep[see also][]{wada02,bournaud10,hopkins11}. The relationship they derive between \sfrsd{} and gas velocity dispersion ($\sigma_{g}$) is given by their equation (59), which assumes MKS units:

\begin{multline} \label{eq:k18}
    \Sigma_{\mathrm{SFR}} = f_{\mathrm{sf}} \frac{\sqrt{8(1+\beta)} f_{g,Q}}{GQ} \frac{\sigma_{\mathrm{g}}}{t_{\mathrm{orb}}^{2}} \\ \times \mathrm{max}\left[\frac{8 \epsilon_{\mathrm{ff}} f_{g,Q}}{Q} \sqrt{\frac{2(1+\beta)}{3 f_{g,P} \phi_{\mathrm{mp}}}} , \frac{t_{\mathrm{orb}}}{t_{\mathrm{sf,max}}} \right]
\end{multline}

\noindent In this equation, $f_{\mathrm{sf}} = [1.0,0.5]$ is the fraction of the ISM in the molecular phase, $f_{g,Q} = [0.7,0.5]$ is the fractional contribution of gas to $Q$, $t_{\mathrm{orb}} = [200,200]$~Myr is the galaxy orbital time, $f_{g,P} = [0.7,0.5]$ is the fractional contribution of the gas self-gravity to the midplane pressure; these values are for high-redshift and local spiral galaxies respectively. The remaining parameters are the same for both high-redshift galaxies and local spirals: $\beta = 0$ is the rotation curve index, $G$ is the gravitational constant, $Q = 1$ is the Toomre $Q$ parameter, $\epsilon_{\mathrm{ff}} = 0.015$ is the star-forming efficiency per free-fall time, $\phi_{\mathrm{mp}} = 1.4$ is the ratio of total pressure to turbulent pressure at the midplane, and $t_{\mathrm{sf,max}} = 2$~Gyr is the maximum star-forming timescale. The values we quote here are the fiducial values adopted by \citet[][see their Table 1 and 3]{krumholz18}; using these values and converting to MKS units, we derive the following expression for the \smol{}$-$\sfrsd{} relation:

\begin{multline} \label{eq:k18_smol}
    \sigma_{\mathrm{mol,high-}z} = \frac{1}{460 \times 0.977} \, t_{\mathrm{orb}}^{2} \Sigma_{\mathrm{SFR}} \\ \times \mathrm{max}\left[\frac{8 \epsilon_{\mathrm{ff}} f_{g,Q}}{Q} \sqrt{\frac{2(1+\beta)}{3 f_{g,P} \phi_{\mathrm{mp}}}} , \frac{t_{\mathrm{orb}}}{t_{\mathrm{sf,max}}} \right]^{-1}
\end{multline}

\noindent for high-redshift galaxies (yellow dashed curve in Figure \ref{fig:sh2_comp}), and: 

\begin{equation}
    \sigma_{\mathrm{mol,spiral}} = 2.8 \times \sigma_{\mathrm{mol,high-}z}
\end{equation}

\noindent for local spirals (yellow dotted curve in Figure \ref{fig:sh2_comp}). This framework predicts that the primary mechanism driving turbulence is a function of mass and redshift, with gravity-driven turbulence dominating in high-redshift galaxies and high masses, and feedback-driven turbulence dominating at lower redshifts and masses. Furthermore, the model predicts that the maximum velocity dispersion that can be maintained by feedback alone is ${\sim} 6-10$~\kms{}, where the exact value depends on the gas fraction, thermal velocity dispersion, and fraction of the ISM that is in the molecular hydrogen phase. 

In the relevant properties for this theory (e.g., gas fraction, Toomre $Q$; see Section \ref{subsec:sample}), DYNAMO galaxies are most similar to those of $z\sim1$ galaxies. There is evolution of properties such as the \sfrsd{} and velocity dispersion from $z\sim1$ to $z\sim2$; therefore we expect that DYNAMO galaxies may fall between the ``spiral" and ``high$-z$" categories from \cite{krumholz18}.

However, our comparison reveals that for the assumed fiducial parameters, both the local spiral and high-redshift curves overpredict the velocity dispersions that we observe in DYNAMO and those of PHANGS galaxies. This observation was also made in \citet{girard21} for global DYNAMO measurements of \smol{} and SFR (see their Figure 4) and in \citet{Roman-Oliveira2024} for a sample of four $z \sim 4.5$ galaxies (see their Figure 5).

\begin{figure}
    \centering
    \includegraphics[width=\columnwidth]{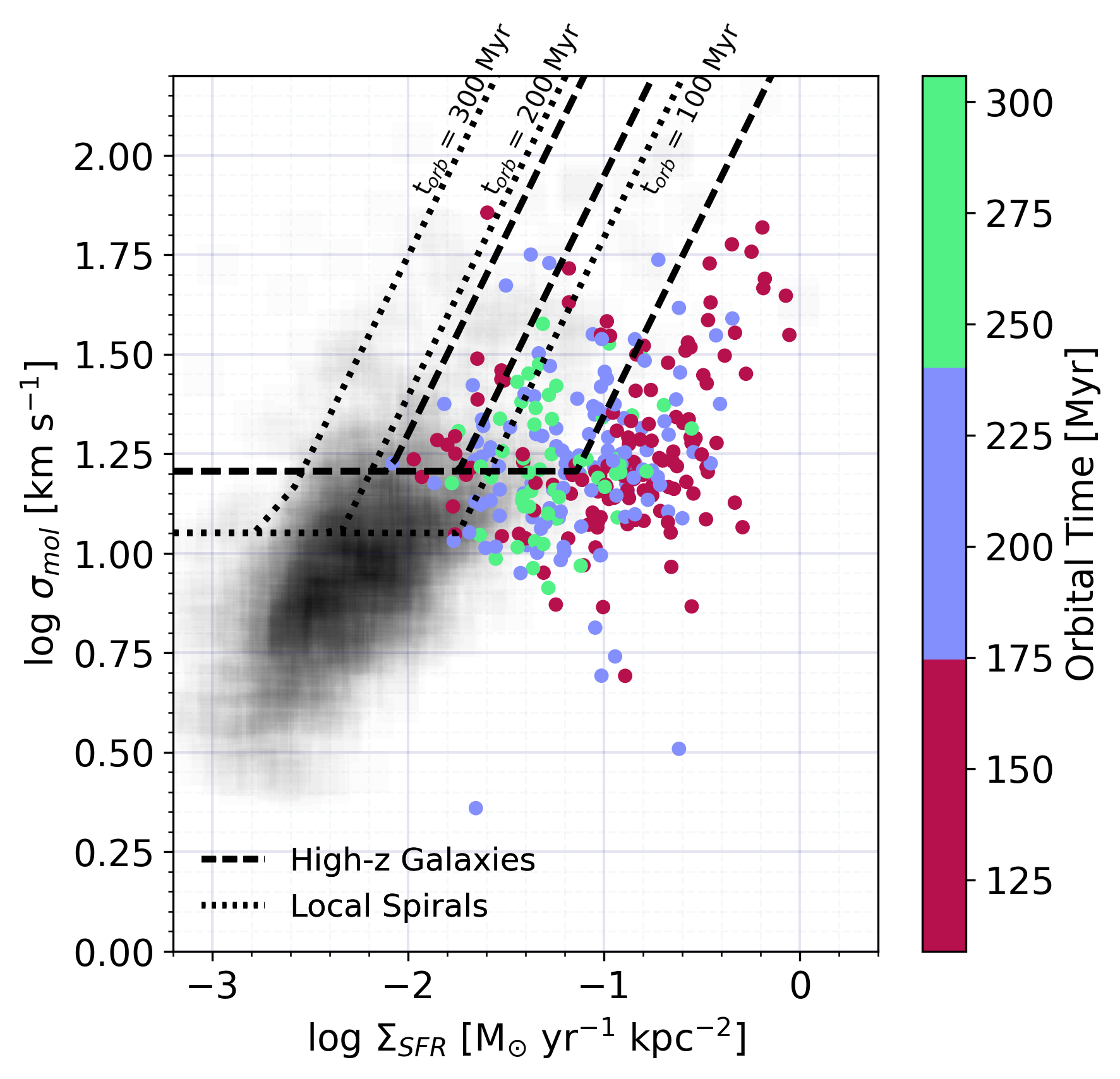}
    \caption{The same as the left panel of Figure~\ref{fig:sh2_comp}; however, we now only include DYNAMO measurements at larger radii where the rotation curves are flat and the $\beta = 0$ assumption is valid. The range of \torb{} for the DYNAMO points in this case is ${\sim} 100-300$~Myr. The black dashed (dotted) lines are the \citet{krumholz18} models for the high$-z$ (local spiral) case, assuming $t_{\mathrm{orb}} = 300,200,100$~Myr (from left to right). Because DYNAMO galaxies are most similar to $z \sim 1$ galaxies, we expect them to lie between the dotted and dashed model lines for this range of \torb{}. However, we find that while the shape of the transport+feedback models qualitatively match DYNAMO, the observed velocity dispersions are still over-predicted.}
    \label{fig:k18_torb}
\end{figure}

The break in the model of equation~(\ref{eq:k18_smol}) can be shifted to higher and lower \sfrsd{} values by revisiting the values we assume for the fiducial parameters: (1) decreasing the fraction of the ISM that is in the molecular phase, $f_{\mathrm{sf}}$, shifts the break in the model to smaller \sfrsd{} values which increases the discrepancy with the DYNAMO observations, (2) increasing the fractional contribution of the gas to Q, $f_{\mathrm{g,Q}}$, to 1 shifts the model break to only marginally higher \sfrsd{}, and (3) decreasing the orbital time, \torb{}, shifts the break to higher \sfrsd{} values. To further explore this last option, we replicate the left panel of Figure~\ref{fig:sh2_comp} in Figure~\ref{fig:k18_torb}. First, we compute the rotation curve index, $\beta = d \, \mathrm{ln (v)} / d\, \mathrm{ln}(r)$, for each galaxy. We then determine the radius at which $\beta$ deviates from a flat rotation curve ($\beta = 0$) by 25\%. Because we are assuming $\beta = 0$ in equation \ref{eq:k18}, for this comparison we exclude in Figure \ref{fig:k18_torb} any DYNAMO data points where the rotation curve is rising and $\beta \neq 0$. We now also color the DYNAMO data points according to the \torb{} value at the corresponding radius ($t_{\mathrm{orb}} = 2\pi R/V$ and correcting for inclination). Where the rotation curves are flat, we see that the range of \torb{} values for DYNAMO is ${\sim} 100-300$~Myr. Thus, we now overplot the \citet{krumholz18} high$-z$ (black dashed lines) and local spiral (black dotted lines) models for $t_{\mathrm{orb}} = 300, 200, 100$~Myr, from left to right. As DYNAMO galaxies are most similar to star-forming systems at $z \sim 1$, we expect the data points to fall somewhere between the high$-z$ and local spiral lines. Instead, we find that the models overpredict the observed velocity dispersion measurements as the majority of the DYNAMO points fall to the right of the high$-z$, $t_{\mathrm{orb}} = 100$~Myr model. We note however that the general shape of the transport+feedback model provides a qualitative match to our observations.


Studies of global star forming and molecular gas properties in high-redshift galaxies find similar results. \citet{Roman-Oliveira2024} compare global gas velocity dispersions from ALMA \CII{} observations and SFRs inferred from total-infrared luminosity measurements for four $z \sim 4.5$ discs to the transport and feedback models of \citet{krumholz18}. They find that for any assumption of maximum circular speed \citep[][see equation (60)]{krumholz18} of the galaxies, the transport models over-predict the observations (see their Figure 5, left panel) consistent with our results, while the feedback only models (see their Figure 5, right panel) do not. \citet{Rizzo2024} find similar results for a sample of 57 $z = 0-5$ galaxies where cold gas tracers (CO, \CI{}, \CII{}) are used to measure global velocity dispersions (see their Figure 4).

Observations show that galaxies must be continuously replenished with gas to maintain star formation on timescales longer than the typical $1-2$~Gyr gas depletion times in local galaxies or the even shorter depletion times of high-redshift galaxies. Simulations show that cold, smooth streams of gas join the disks of high-redshift galaxies at large radii \citep[${\sim} 0.1-0.3 \times$ the virial radius, or ${\sim} 10 \times$ the stellar scale length;][]{danovich15}, generating large gas surface densities in these outer regions \citep[see also][]{keres05,dekel09,trapp22}. Motivated by this, \citet{ginzburg22} built upon the work of \citet{krumholz18} by including the conversion of accretion energy into turbulent energy as a third mechanism for driving turbulence in disks \citep[see also][]{elmegreen10,klessen10}. They find that galaxies in dark matter halos that evolve to masses ${\leq} 10^{12}$~M$_{\odot}$ by redshift $z = 0$ are dominated by feedback-driven turbulence. For galaxies in more massive halos, they are dominated by transport-driven turbulence or accretion-driven turbulence depending on the efficiency of converting accretion kinetic energy into turbulent energy. However, the effect of adding accretion-driven turbulence is to increase the floor in \smol{} that can be maintained by stellar feedback alone, which increases the discrepancy between our observations. It also decreases the level of star formation required to achieve very high velocity dispersions, which also increases the discrepancy with our observations \citep[see Figure 6 in][]{ginzburg22}.

Finally, we compare our \smol{}$-$\gsd{} relation to the work of \citet{brucy20}, who conduct magnetohydrodynamic simulations of 1~kpc-sized cubic regions in local and high$-z$, gas-rich disks where the effects of stellar feedback (\HII{} region formation, supernovae, and far-ultraviolet feedback) and turbulent forcing on large scales by an external force are considered independently and together. The authors find that stellar feedback alone is enough to suppress star formation to levels of the KS relation in nearby galaxies, but is insufficient to do so in high$-z$ systems. For these, the authors argue that large-scale turbulent driving, either from mass accretion onto the galaxy, gas transport within the disk, or some other mechanism, is required \citep[see also][]{brucy23}. They test two scalings for the turbulent driving: one where the mean power injected is $\overline{P_{\mathrm{inj}}} \propto \Sigma_{0,\mathrm{gas}}^{2.5}$, and a stronger form where $\overline{P_{\mathrm{inj}}} \propto \Sigma_{0,\mathrm{gas}}^{3.8}$ ($\Sigma_{0,\mathrm{gas}}$ is the initial gas surface density in each simulation). The authors find that the stronger form of turbulent driving produces a KS law slope more consistent with observations. In addition, they measure the gas velocity dispersion as a function of gas surface density (see their Figure A1). Although the range of gas surface densities probed is small ($10-100$~M$_{\odot}$\,pc$^{-2}$), the feedback only simulation (large-scale turbulent driving turned off) produces small velocity dispersions incompatible with our observations. The weaker turbulent driving simulation ($\overline{P_{\mathrm{inj}}} \propto \Sigma_{0,\mathrm{gas}}^{2.5}$) with stellar feedback produces the relation: 
\begin{equation}
    \mathrm{log} \, \sigma_{\mathrm{mol}} = 0.65 \, \times \, \mathrm{log} \, \Sigma_{\mathrm{mol}} + 0.41
\end{equation}

\noindent which is steeper than the best-fit slope of $N = 0.47$ that we find (see the left panel of Figure \ref{fig:disp_sh2}). Due to the scatter in the observational measurements, this relation is not inconsistent with the DYNAMO+PHANGS data. However, this weaker turbulent driving simulation produces a steeper KS-relation slope than is observed. The stronger turbulent driving simulation ($\overline{P_{\mathrm{inj}}} \propto \Sigma_{0,\mathrm{gas}}^{3.8}$) produces a much steeper relation with a slope of 1.2. 

\subsubsection{Feedback-Regulated Star Formation}
In contrast, feedback-regulated models of star formation, which we compare to in the right panel of Figure \ref{fig:sh2_comp}, argue that the primary source of energy in the disk of galaxies is young stars, and the feedback from these stars balances the weight of the ISM \citep[also referred to as pressure-regulated feedback-modulated star formation, PRFM;][]{ostriker22}. Turbulent pressure is maintained by energy/momentum injected from supernovae, and is proportional to the star formation rate per area. Thermal pressure is maintained by photoelectric heating of the gas by stellar far-ultraviolet (FUV) photons and is also proportional to the star formation rate. The disk is then maintained in a state of quasi-equilibrium through a self-regulation of the SFR, such that the energy injection by stellar feedback balances the dissipation of turbulent energy and the cooling of the ISM \citep[see][for more theory details, numerical simulations, and applications]{ostriker10,kim11,ostriker11,shetty12,kim13,faucher-giguere13,kim15,hayward17,orr18,gurvich20}. 

\citet{ostriker22} revisit this theory and estimate the turbulent, thermal, and magnetic pressures to predict the total feedback yield ($\Upsilon_{\mathrm{tot}} = P_{\mathrm{tot}}/\Sigma_{\mathrm{SFR}}$). They then conduct magnetohydrodynamic simulations within the ``Three-phase Interstellar Medium in Galaxies Resolving Evolution with Star Formation and Supernova Feedback'' \citep[TIGRESS;][]{kim17} numerical framework to evaluate $\Upsilon_{\mathrm{tot}}$ and test the predictions of PRFM theory. The authors use seven TIGRESS simulations which model a three-phase ISM with varying initial gas surface densities in $512 \times 512$~pc$^{2}$ to $2048 \times 2048$~pc$^{2}$ galaxy patches and a vertical dimension that is $7\times$ as large. The stellar feedback mechanisms included are supernova explosions and the effects of FUV radiation \citep[see][for more details]{kim20a}. The authors measure in their simulations feedback yields that are consistent with their theoretical predictions and a \sfrsd{}$-P_{\mathrm{DE}}$ relation that is consistent with observations (where $P_{\mathrm{DE}}$ is the dynamical equilibrium pressure, an estimate of the ISM weight; see their Figure 15).

To compare our measurements to the results of \citet{ostriker22}, we derive an expression relating the gas velocity dispersion to \sfrsd{} by combining their equation (28):
\begin{equation}
    \frac{\Sigma_{\mathrm{SFR}}}{\mathrm{M_{\odot}\,pc^{-2}\,Myr^{-1}}} = 2.07 \times 10^{-4} \; \frac{P_{\mathrm{DE}}/k_{B} \, [\mathrm{cm^{-3}\,K}]}{\Upsilon_{\mathrm{tot}} \, [\mathrm{km\,s^{-1}}]}
\end{equation}

\noindent with an expression for velocity dispersion in their Section 4.6:
\begin{equation}
    \sigma = 12 \, [\mathrm{km\,s^{-1}}] \left(\frac{P_{\mathrm{DE}}}{10^{4} \, k_{B} \, [\mathrm{cm^{-3}\,K}]}\right)^{0.22}
\end{equation}

\noindent and express the velocity dispersion as a function of \sfrsd{} and $\Upsilon_{\mathrm{tot}}$, the total feedback yield (ratio of the total pressure to \sfrsd{}):
\begin{equation} \label{eq:o22}
    \sigma = 12 \, [\mathrm{km\,s^{-1}}] \left(\frac{1}{2.07} \Sigma_{\mathrm{SFR}} \Upsilon_{\mathrm{tot}}\right)^{0.22}
\end{equation}

\noindent where $\Upsilon_{\mathrm{tot}}$ is given by:
\begin{equation}
    \Upsilon_{\mathrm{tot}} = 740 \, [\mathrm{km\,s^{-1}}] \left(\frac{\Sigma_{\mathrm{SFR}}}{0.01 \, [\mathrm{M_{\odot}\,pc^{-2}\,Myr^{-1}}]}\right)^{-0.18}.
\end{equation}

\noindent Combining these two equations results in a \smol{}$-$\sfrsd{} relation of:

\begin{equation}
    \mathrm{log} \, \sigma_{\mathrm{mol}} = 0.1804 \, \times \, \mathrm{log} \, \Sigma_{\mathrm{SFR}} + 1.56
    \label{eq:log_o22}
\end{equation}

\noindent which we plot in Figure \ref{fig:sh2_comp} (solid purple line) and provides reasonable agreement with our DYNAMO measurements. The intercept of $C = 1.56$ predicted by this model is well-matched to what we find for our observations, $C = 1.56 \pm 0.02$, while the slope of $N = 0.1804$ is ${\sim } 4.5\sigma$ lower than the derived slope for DYNAMO, $N = 0.27 \pm 0.02$.

In the right panel of Figure \ref{fig:sh2_comp}, we also compare to the numerical results from \citet{Rathjen2023}. This work is built upon the ``Simulating the Life Cycle of Molecular Clouds'' \citep[SILCC;][]{walch15} framework and aims to investigate the effect of cosmic rays on the multiphase structure of star formation-driven outflows. The authors conduct magnetohydrodynamic simulations of the ISM in a $500 \times 500 \pm 4000$~pc$^{3}$ galactic patch, where they model star formation using sink particles and track the evolution of individual massive stars ($8-120$~M$_{\odot}$). Their feedback model includes the effects of core-collapse supernovae, stellar winds, ionizing radiation from massive stars, and cosmic rays. These simulations show that cosmic rays are important for establishing outflows with a cold, warm, and hot component. When measuring the velocity dispersion of the cold neutral medium in all simulated environments with cosmic rays, the authors find a power law relation between \smol{} and \sfrsd{} of the form:
\begin{equation}
    \mathrm{log} \, \sigma_{\mathrm{mol}} = (0.20 \pm 0.02) \, \times \, \mathrm{log} \, \Sigma_{\mathrm{SFR}} + (1.30 \pm 0.02).
    \label{eq:r23}
\end{equation}

\noindent This is what we show in the right panel of Figure \ref{fig:sh2_comp} (purple dashed line) and it also shows good agreement with our resolved DYNAMO measurements. The intercept predicted by this model is much lower than what we find for DYNAMO; as such, the majority of DYNAMO points lie above equation~(\ref{eq:r23}), while the slope of $N = 0.20 \pm 0.02$ is within ${\sim} 2.5\sigma$ of our measured slope of $N = 0.27 \pm 0.02$. This model also provides a very good match to our fit of the global DYNAMO \smol{}$-$\gsd{} relation (see also Figure \ref{fig:globalsig} and equation \ref{eq:global_smol}). 

Finally, we also include as a comparison the stellar feedback-only model of \citet{krumholz18} in the fixed Q and variable $\epsilon_{\mathrm{ff}}$ case (purple dotted line; their equation 61):

\begin{equation}
    \Sigma_{\mathrm{SFR}} = \frac{8 (1+\beta) \pi \eta \sqrt{\phi_{\mathrm{mp}} \phi_{\mathrm{nt}}^{3}} \, \phi_{\mathrm{Q}}}{G Q^{2} \left<p_{*}/m{*}\right> f_{\mathrm{g,P}}} \, \frac{\sigma_{\mathrm{g}}}{t_{\mathrm{orb}}^{2}}
    \label{eq:k18_feedback}
\end{equation}

\noindent where $\eta = 1.5$ is a scaling factor for the turbulent dissipation rate, $\phi_{\mathrm{nt}} = 1$ is the fraction of the velocity dispersion that is non-thermal, $\phi_{\mathrm{Q}} = 2$ is defined as one plus the ratio of the gas to stellar $Q$, and all remaining terms are the same as in equation~(\ref{eq:k18}). We have assumed \torb{}$= 200$~Myr as in the transport+feedback case. Combining this with the additional terms in equation~(\ref{eq:k18_feedback}) results in a \smol{}$-$\sfrsd{} relation of: 

\begin{equation}
    \mathrm{log} \, \sigma_{\mathrm{mol}} = 0.5 \, \times \, \mathrm{log} \, \Sigma_{\mathrm{SFR}} + 1.81
\end{equation}

\noindent where the slope of $N = 0.5$ is 11.5$\sigma$ higher than our best-fit slope. Although the intercept and slope in this case are larger than the best-fit \smol{}$-$\sfrsd{} relation we find, this model passes reasonably well through the DYNAMO points but underestimates the PHANGS velocity dispersions. Assuming a shorter \torb{} shifts the \citet{krumholz18} model down (i.e., smaller normalization) which results in poorer agreement with the data. We find similar rms residuals between these three feedback-only models and our data; however, the power-law slopes of the SILCC \citep{Rathjen2023} and TIGRESS \citep{ostriker22} models are in better agreement with the observed slope measured here.

We conclude from these model comparisons that stellar feedback alone is sufficient to reproduce the \smol{}$-$\sfrsd{} relation we observe at kpc scales in the gas-rich galaxies of our DYNAMO sample and of the local star-forming PHANGS galaxies.


\subsection{Feedback-Driven Outflows} \label{subsec:outflows}
\begin{figure}
    \centering
    \includegraphics[width=\columnwidth]{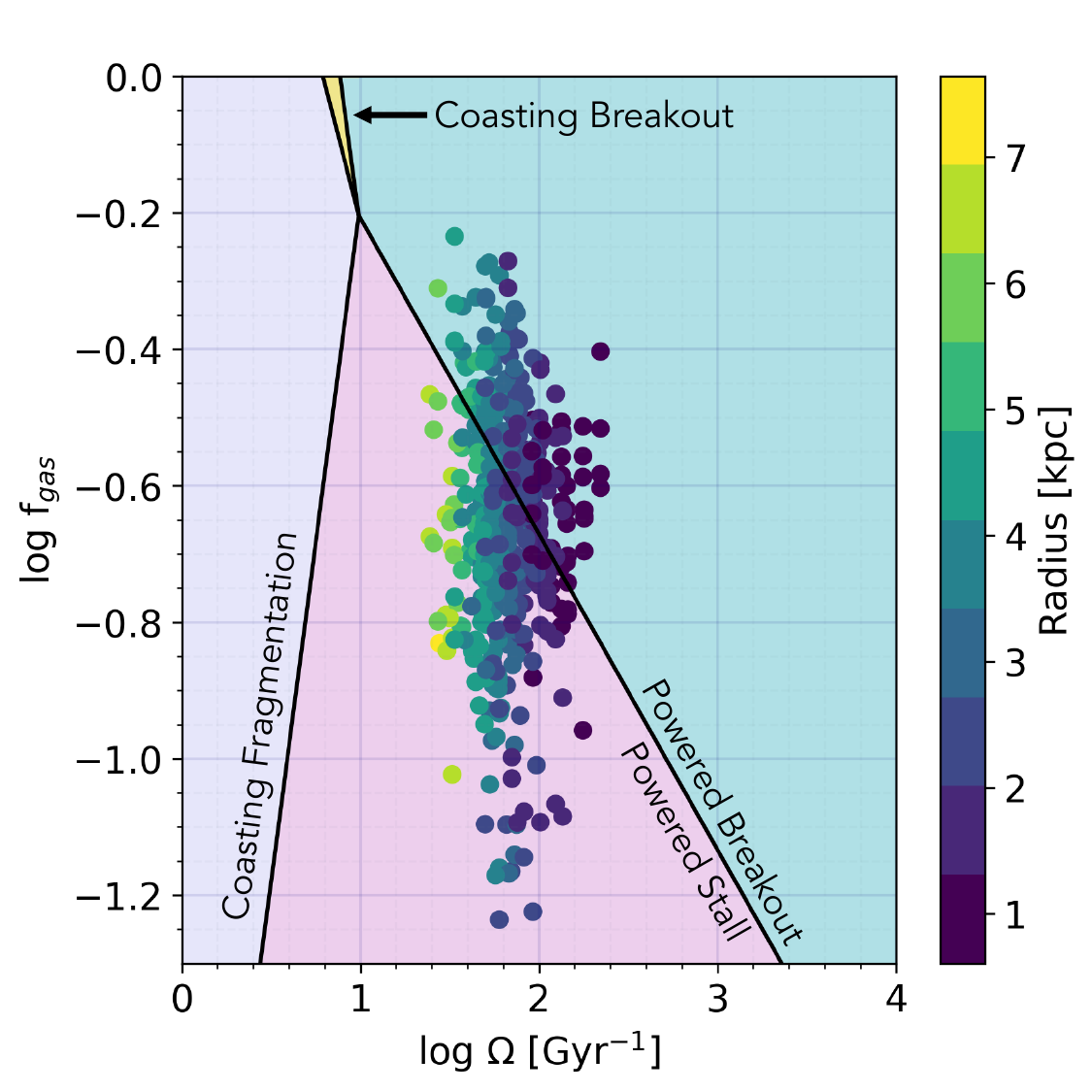}
    \caption{Gas fraction ($f_{\mathrm{gas}}$) as a function of orbital frequency ($\Omega = R/V$). The green shaded region of this parameter space is where superbubbles created by clustered supernovae are expected to reach the scale height of the disk and breakout to produce outflows before the final supernova goes off \citep[``powered breakout'';][]{orr22a,Orr2022b}. The pink shaded region is where the superbubble expansion stalls within the disk before reaching the disk scale height and before the final supernova is produced (``powered stall''). DYNAMO line-of-sight measurements (colored circles) are shaded according to the distance of the region from the galaxy center. We find that ${\sim} 38$\% of our DYNAMO measurements lie in the region where outflows and/or fountains are expected to occur.}
    \label{fig:outflows}
\end{figure}

\citet{orr22a,Orr2022b} develop an analytic model to investigate the effects of temporally and spatially clustered supernovae on star formation regulation and the launching of outflows from disk galaxies. Because massive stars only have a short window within which they can undergo a core-collapse supernova (${\sim} 40$~Myr), the detonation of supernovae is expected to be clustered in space and time and their overlap will create large expanding cavities within the host disk \citep[i.e., a ``superbubble''; see also][]{Fielding2018}. Such structures are now readily observed in nearby galaxies \citep[with JWST; see e.g.,][]{Barnes2023,Watkins2023} and in our own Galactic neighborhood \citep{Zucker2022}. In the analytical model of \citet{Orr2022b}, superbubbles can expand to reach the scale height of the disk and breakout before the last supernova detonates (powered breakout) or after (coasting unpowered breakout).  In the breakout case, more than 60\% of the feedback momentum can be lost to powering outflows and fountains rather than driving turbulence in the ISM, which results in a reduction of $(P_{*}/m_{*})$. Likewise, the superbubble expansion can stall in the ISM before the final supernova (powered stall) or after (coasting unpowered fragmentation). These four scenarios can be distinguished in gas fraction ($f_{\mathrm{gas}}$) versus orbital frequency ($\Omega$) parameter space \citep[see Table 1 in][for boundary equations]{orr22a}.

We plot $f_{\mathrm{gas}}-\Omega$ in Figure \ref{fig:outflows} with our DYNAMO line-of-sight measurements colored according to their distance from their galaxy center. We find that ${\sim} 38$\% of our measurements lie within the region where powered breakouts, and therefore outflows and/or fountains, are expected to occur. Disk locations as far as ${\sim} 6$~kpc from their host galaxy center are found in the powered breakout region, suggesting that outflows may be present, and an important star-formation regulator, within DYNAMO.

\section{Summary and Conclusions} \label{sec:conclusion}
In this work, we have combined ${\sim} 1-2$~kpc scale ALMA observations of \jupthree{} and \jupfour{} with HST to study star-formation laws in gas-rich star-forming disks. Specifically, we use a sample of seven DYNAMO galaxies to investigate the ``molecular gas main sequence'', the KS relation, and the relation of molecular gas velocity dispersion to the SFR and molecular gas mass surface densities which allows us to test theories of star formation regulation. We summarize our findings here: \\

\noindent 1. DYNAMO galaxies are more gas-rich than local spirals (see Figure \ref{fig:ssd_gsd}) and are not analogous to the centers of local star-forming galaxies. Rather, they lie above the molecular gas main sequence derived from measurements of nearby systems (e.g., EDGE, ALMaQUEST, PHANGS). Our DYNAMO measurements are fit by a shallower slope and larger normalization than what has been found in previous studies of local galaxies. \\

\noindent 2. The resolved \gsd{}$-$\sfrsd{} relation in DYNAMO galaxies (see Figure \ref{fig:ks_law}) has a near linear slope of $N = 0.90 \pm 0.04$ when assuming a fixed, Milky Way-like \aco{} factor. Under this assumption, more than half of the line-of-sight measurements across our sample have molecular gas depletion times shorter than 1~Gyr. This is in contrast to local galaxies, such as PHANGS and HERACLES, which have a median depletion times of ${\sim} 2$~Gyr. Adopting a variable \aco{} \citep[equation 31 in][]{bolatto13} results in a steeper slope of $N = 1.23 \pm 0.03$. In this case, nearly all positions across the seven DYNAMO disks have depletion times shorter than 1~Gyr. \\

\noindent 3. Compared to local galaxies from PHANGS, DYNAMO galaxies have high SFR surface densities, gas surface densities, and velocity dispersions throughout their disks, not just their centers (see Figure \ref{fig:disp_sh2}). \\

\noindent 4. We compare our \smol{}$-$\sfrsd{} relation to predictions from star formation regulation theories (see Figure \ref{fig:sh2_comp}) that incorporate stellar feedback and gravitational instabilities as mechanisms for driving turbulence \citep[e.g.,][]{krumholz18}, and feedback-regulated star formation theories that invoke stellar feedback alone \citep[e.g.,][]{ostriker22}. For the fiducial parameters adopted in \citet{krumholz18}, we find that the gravity+feedback model overpredicts \smol{} when compared to our observations. If we exclude data points where the rotation curves are rising ($\beta \neq 0$), the data samples $t_{\mathrm{orb}} = 100-300$~Myr. For these values of \torb{}, the models still overpredict the observed \smol{} (see Figure \ref{fig:k18_torb}). Similarly, the large-scale turbulent driving included in the simulations of \citet{brucy20} produces a \smol{}$-$\gsd{} relation that is steeper than what we observe. In contrast, the magnetohydrodynamic simulations of \citet{ostriker22} and \citet{Rathjen2023} predict a relation of \smol{} $\propto$ \sfrsd{}$^{\sim 0.2}$ that provide reasonable matches to both the DYNAMO and PHANGS observations. Finally, the feedback-only \citet{krumholz18} model predicts \smol{} $\propto$ \sfrsd{}$^{0.5}$ and matches the data reasonably well. \\


\noindent 5. Within the $f_{\mathrm{gas}} - \Omega$ parameter space, ${\sim} 38$\% of DYNAMO measurements reside in the region where superbubbles resulting from clustered supernovae may breakout of the disk and drive outflows and/or fountains \citep[see Figure \ref{fig:outflows};][]{orr22a,Orr2022b}. \\

We conclude that the feedback-regulated models of star formation implemented within the TIGRESS \citep{ostriker22} and SILCC \citep{Rathjen2023} magnetohydrodynamic simulations reproduce our observations without the need to invoke additional mechanisms for generating turbulence in the ISM such as gas transport or accretion. Both simulation suites 1) model a galactic patch on ${\sim} 0.5-1$~kpc$^{2}$ scales with a vertical dimension that extends several kpc and physical resolutions of ${\sim} 2-8$~pc, and 2) include the effects of supernovae and FUV radiation in their stellar feedback models. SILCC also includes stellar winds and cosmic rays, while TIGRESS models galactic patches within a differentially rotating disk. However, both produce multi-phase outflows which are important in regulating star formation and producing results in agreement with observations. While outflows have not been observed in this sample of DYNAMO galaxies, our comparison to the analytical work of \citet{Orr2022b} suggests they may be present. JWST can be used to map ionized gas tracers in DYNAMO at high resolution, and modeling of the line profiles can be done to search for ionized gas outflows \citep[see e.g.,][]{ReichardtChu2022a,ReichardtChu2022}.

\acknowledgments
We thank the anonymous referee whose comments have significantly improved this work. 

This paper makes use of the following ALMA data: ADS/JAO.ALMA\#2017.1.00239.S. and ADS/JAO/ALMA\#2019.1.00447.S. ALMA is a partnership of ESO (representing its member states), NSF (USA) and NINS (Japan), together with NRC (Canada), MOST and ASIAA (Taiwan), and KASI (Republic of Korea), in cooperation with the Republic of Chile. The Joint ALMA Observatory is operated by ESO, AUI/NRAO and NAOJ. The National Radio Astronomy Observatory is a facility of the National Science Foundation operated under cooperative agreement by Associated Universities, Inc. Some of the data presented in this article were obtained from the Mikulski Archive for Space Telescopes (MAST) at the Space Telescope Science Institute. The specific observations analyzed can be accessed via \dataset[10.17909/6dmc-ye91]{https://doi.org/10.17909/6dmc-ye91}. 

L.L. acknowledges support from the NSF through grant 2054178 and acknowledges that a portion of their research was carried out at the Jet Propulsion Laboratory, California Institute of Technology, under a contract with the National Aeronautics and Space Administration (80NM0018D0004).

Parts of this research were supported by the Australian Research Council Centre of Excellence for All Sky Astrophysics in 3 Dimensions (ASTRO 3D), through project number CE170100013. 

D.B.F. acknowledges support from Australian Research Council (ARC) Future Fellowship FT170100376 and ARC Discovery Program grant DP130101460.

A.D.B. acknowledges support from the NSF under award AST-2108140.

R.C.L. acknowledges support for this work provided by a NSF Astronomy and Astrophysics Postdoctoral Fellowship under award AST-2102625. 

J.S. acknowledges support by NASA through the NASA Hubble Fellowship grant HST-HF2-51544 awarded by the Space Telescope Science Institute (STScI), which is operated by the Association of Universities for Research in Astronomy, Inc., under contract NAS~5-26555.

D.O. is a recipient of an Australian Research Council Future Fellowship (FT190100083) funded by the Australian Government.

%

\vspace{5mm}
\facilities{ALMA, HST(ACS,WFC)}


\software{aplpy \citep{aplpy2019},
          astropy \citep{astropy13,2018Astropy,2022Astropy},
          \casa\ \citep{mcmullin07},
          numpy \citep{2020NumPy-Array},
          reproject (\href{https://reproject.readthedocs.io/en/stable/}{https://reproject.readthedocs.io/en/stable/},
          SciPy \citep{2020SciPy-NMeth},
          spectral-cube \citep{2019SpectralCube}
          }



\newpage
\appendix
\restartappendixnumbering
\section{Beam Smearing Correction Discussion}
\begin{figure*}
    \centering
    \includegraphics[width=\textwidth]{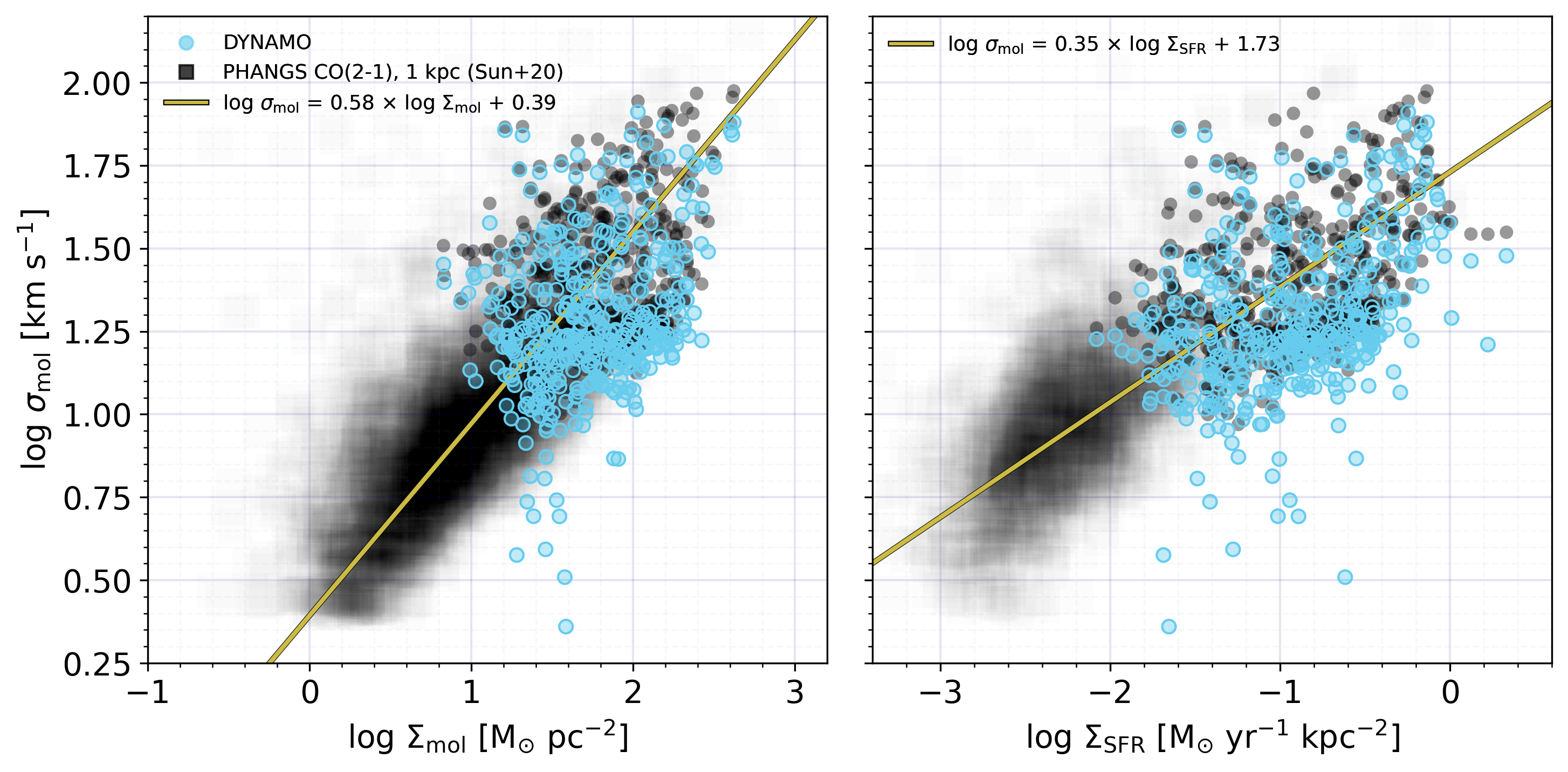}
    \caption{The same as Figure \ref{fig:globalsig}; however, we now include the DYNAMO non-beam smearing corrected velocity dispersion measurements as gray circles. In this case, we fit the power law to these uncorrected data, and find steeper slopes and best-fit relations that are inconsistent with the best-fit relations presented in Table \ref{tab:obs}.}
    \label{fig:uncorr}
\end{figure*}

To assess the effects of our beam smearing correction discussed in section \ref{subsec:gas_disp} on our results, we reproduce Figure \ref{fig:disp_sh2} in Figure \ref{fig:uncorr}, where the gray circles correspond to all velocity dispersion measurements with no beam smearing correction applied. The yellow best-fit lines in both panels now correspond to the uncorrected velocity dispersion measurements. In this case, we find a best-fit slope of $N = 0.58 \pm 0.03$ (versus $N = 0.48 \pm 0.02$) for the \smol{}$-$\gsd{} relation, and $N = 0.35 \pm 0.05$ (versus $N = 0.27 \pm 0.02$) for the \smol{}$-$\sfrsd{} relation.

\begin{figure*}
    \centering
    \includegraphics[width=\textwidth]{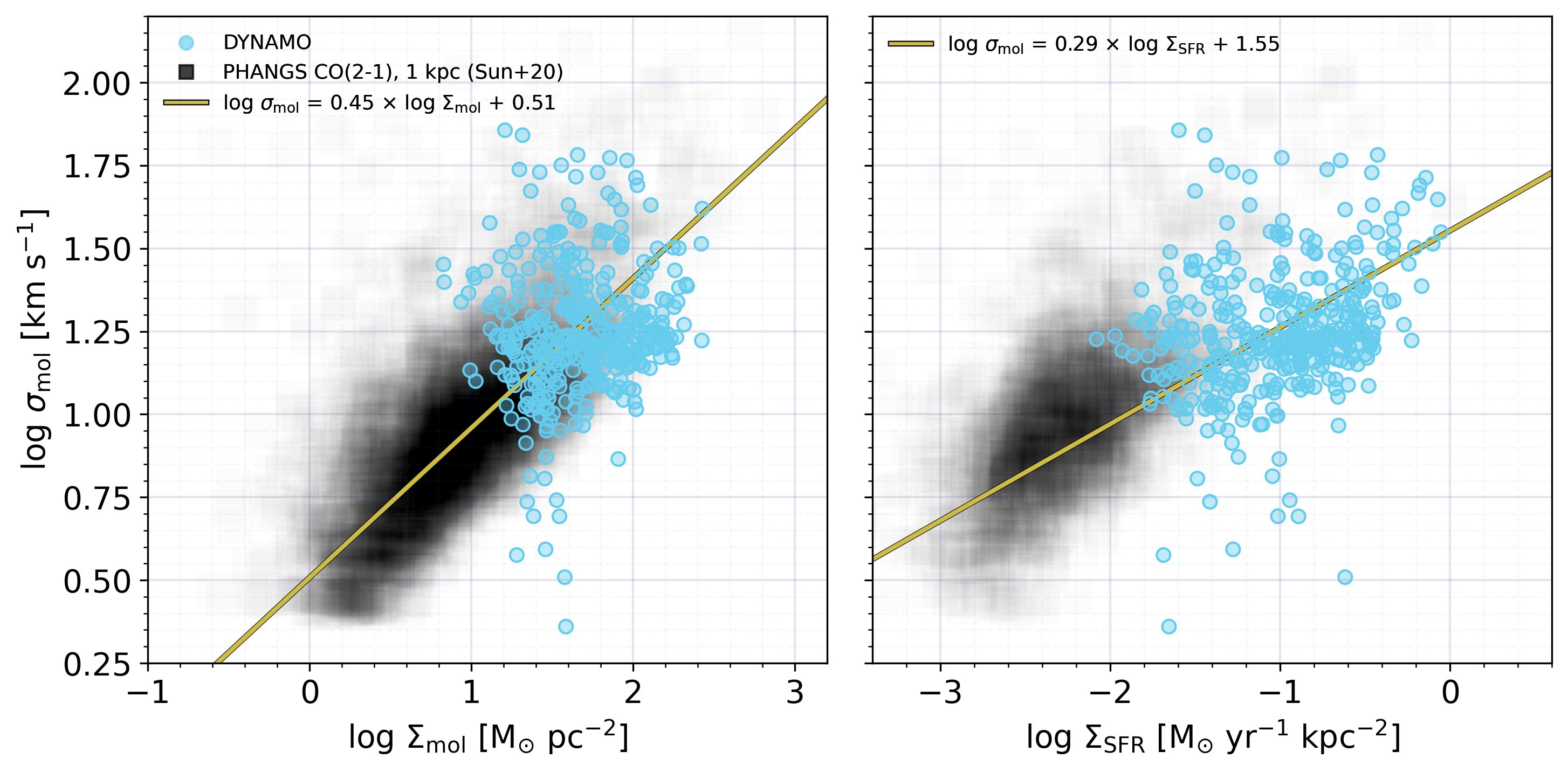}
    \caption{The same as Figure \ref{fig:globalsig}; however, we now exclude all beam smearing corrected measurements that are within two beams of each galaxy center. When we re-fit the power laws to this subset of measurements and find results that are consistent with what is presented in Table \ref{tab:obs}.}
    \label{fig:ex_cen}
\end{figure*}

As an additional test, we again reproduce Figure \ref{fig:disp_sh2} in Figure \ref{fig:ex_cen}; however, we now exclude all velocity dispersion measurements that are within two beams of each galaxy center. Refitting the power laws to this subset of our data, we find best-fit relations that are consistent with what we present in Section \ref{subsec:vdsip_molsd}. Therefore, excluding regions where beam smearing will have the greatest impact on the velocity dispersion measurements does not affect or change the results and conclusions.

\begin{figure*}
    \centering
    \includegraphics[width=0.3\textwidth]{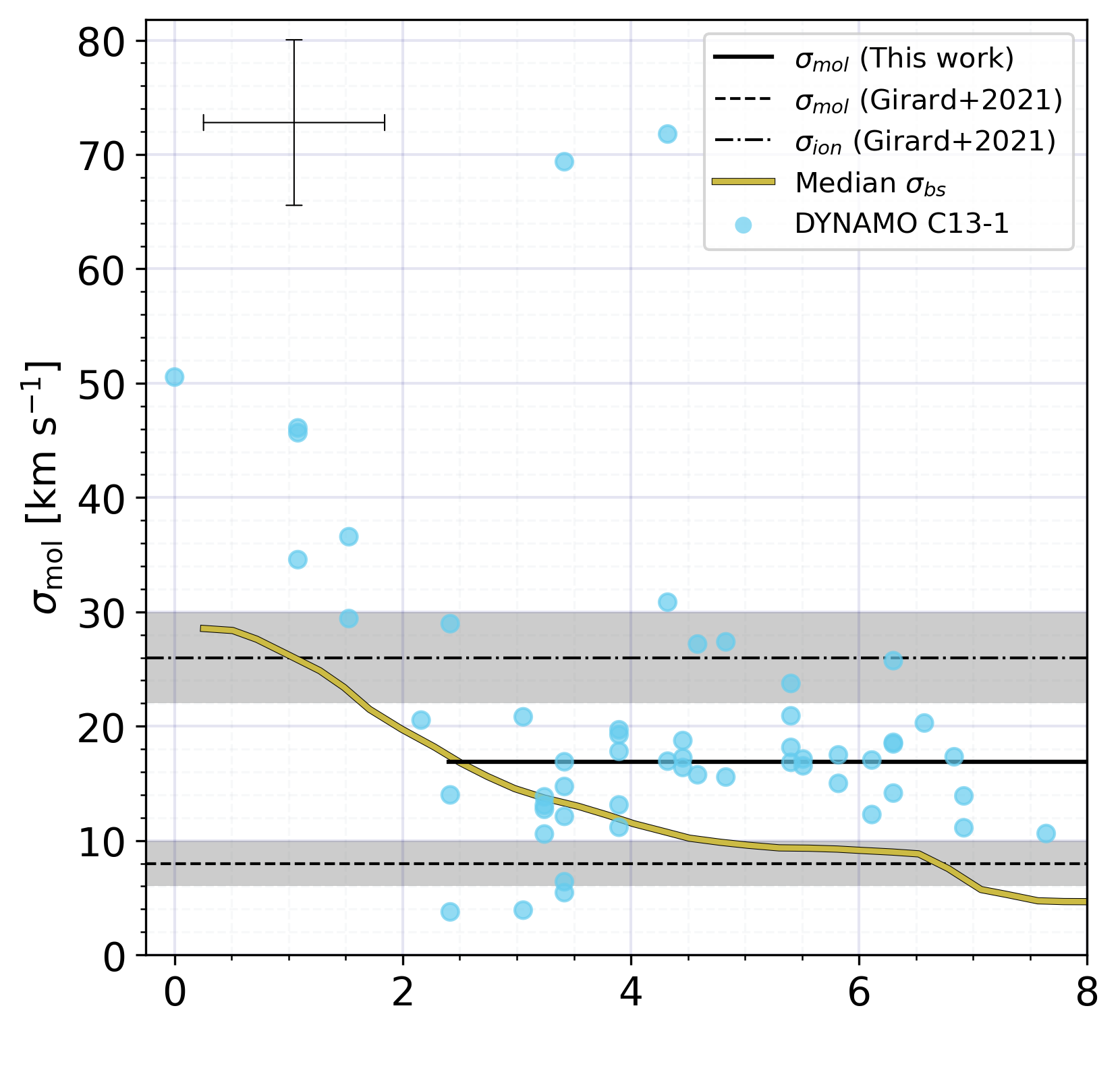}
    \includegraphics[width=0.3\textwidth]{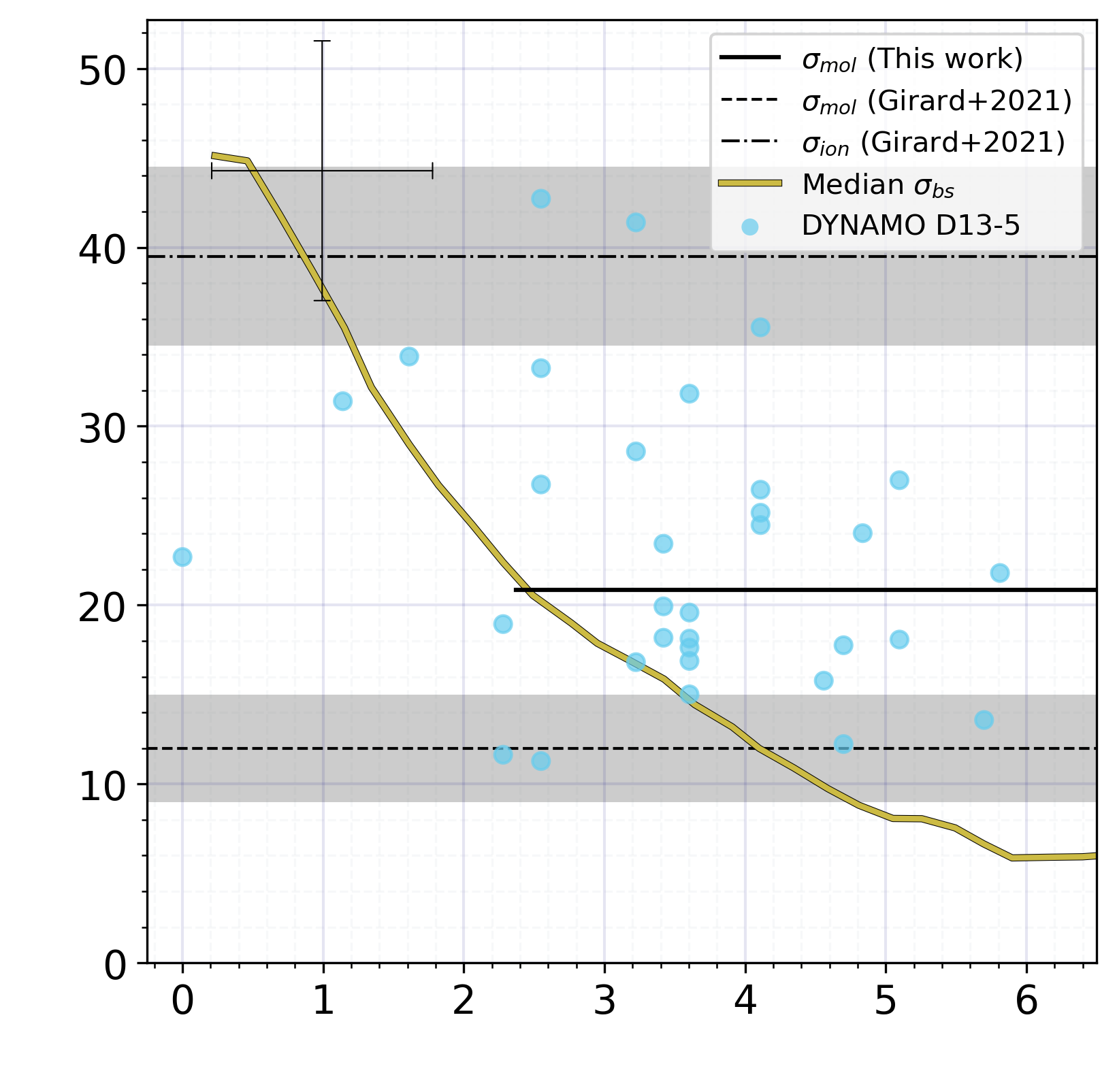}
    \includegraphics[width=0.3\textwidth]{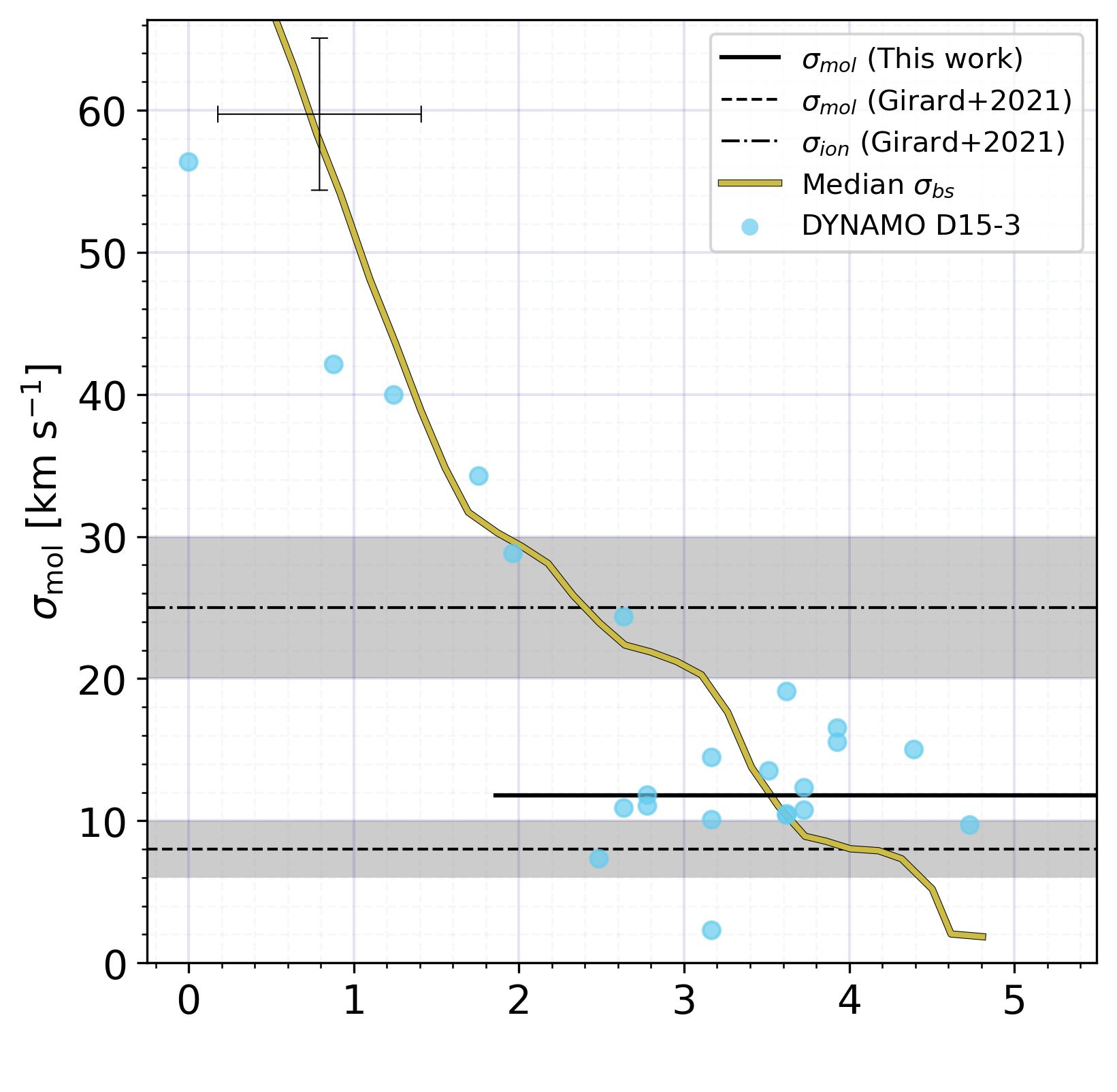}
    
    \includegraphics[width=0.3\textwidth]{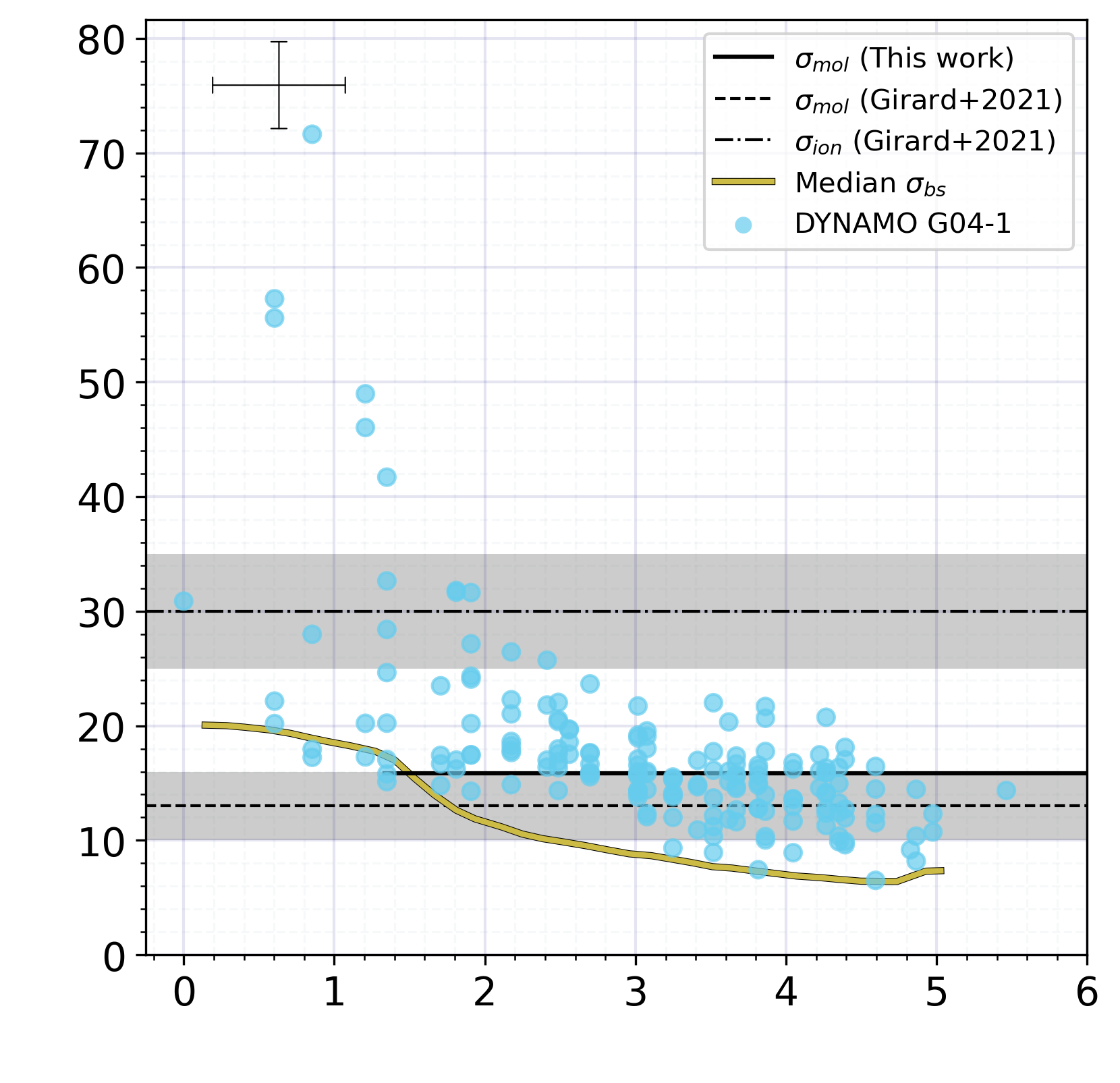}
    \includegraphics[width=0.3\textwidth]{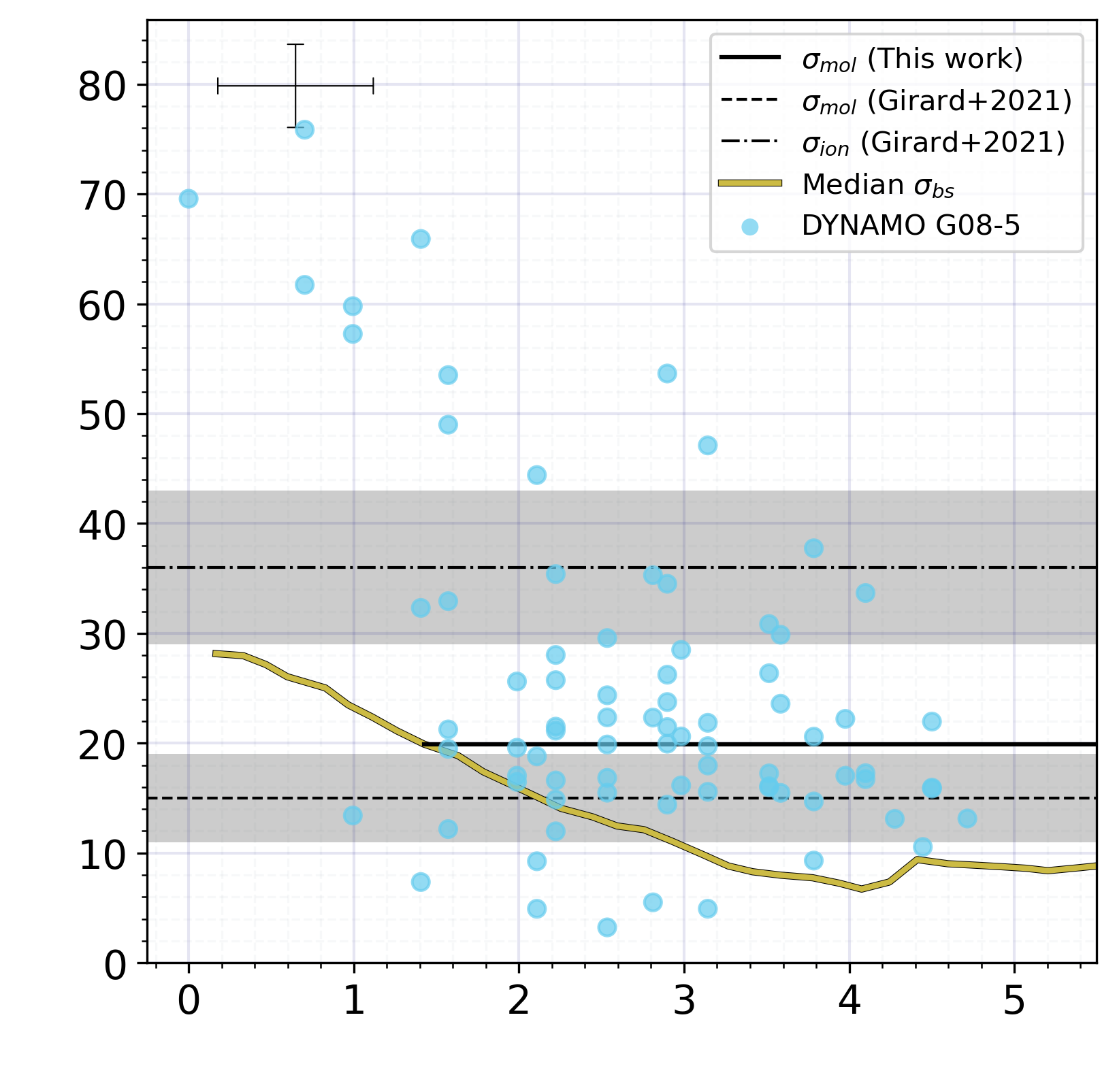}
    \includegraphics[width=0.3\textwidth]{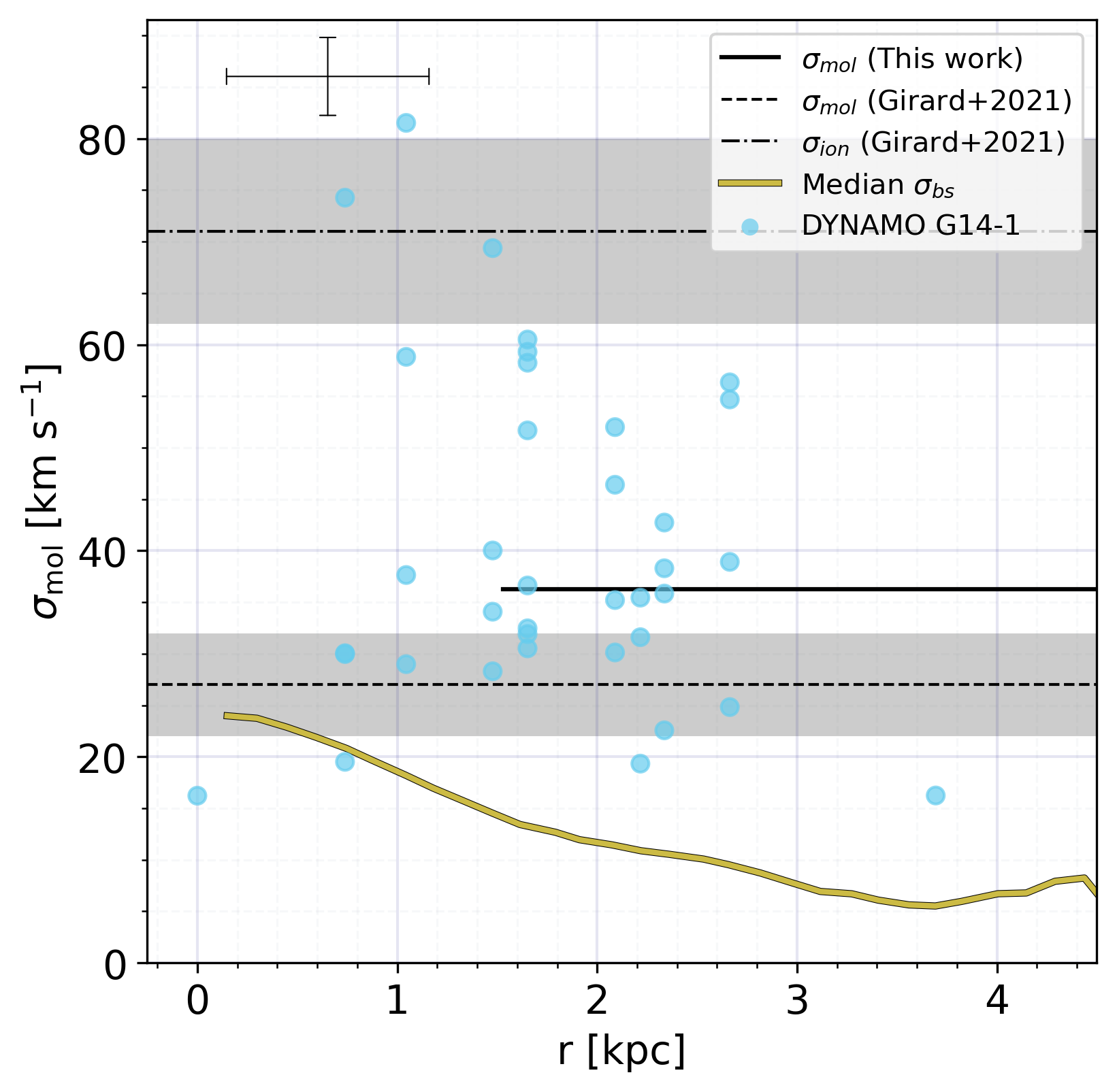}
    
    \includegraphics[width=0.3\textwidth]{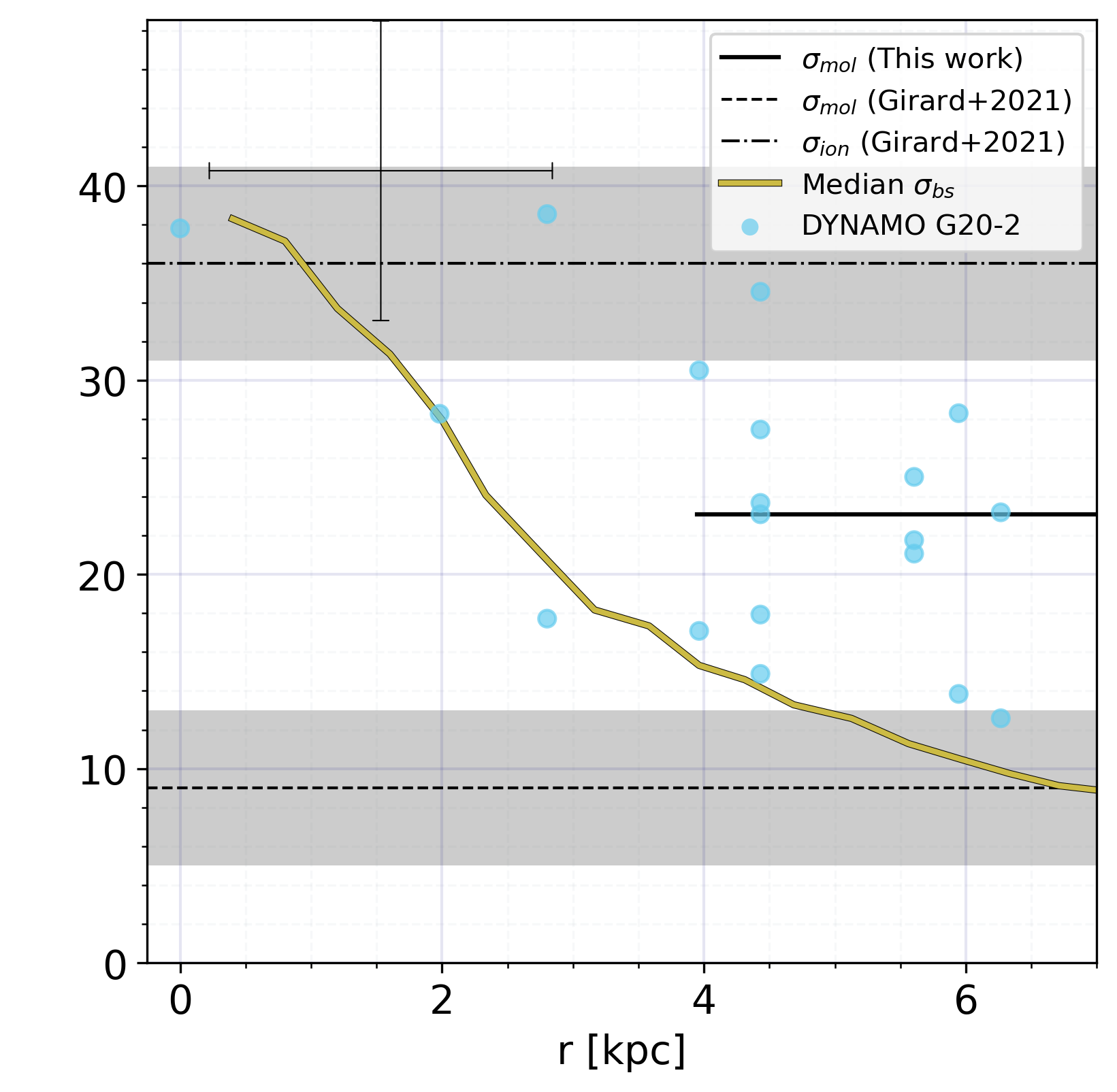}
    \caption{Radial distribution of beam smearing corrected velocity dispersion measurements in DYNAMO along beam-sized sightlines (blue data points). The black solid line corresponds to the median velocity dispersion in DYNAMO galaxies at radii larger than 1.5\,$\times$ the beam FWHM. The error bars in the top left corner of each panel indicates the beam FWHM and the channel size of the \jupthree{} observations used to measure velocity dispersions. For comparison, the black dashed and dot-dashed lines mark the molecular gas and ionized gas velocity dispersion respectively from \citet{girard21}, while the gray shaded region indicates their uncertainties. Within the channel width size, our beam smearing corrected velocity dispersions are consistent with those found by \citet{girard21}. Finally, the yellow solid line is the radial profile of the median beam smearing correction we apply to each velocity dispersion measurement.} 
    \label{fig:rad_disp}
\end{figure*}

Finally, in Figure \ref{fig:rad_disp}, we present the results of our velocity dispersion measurements, corrected for beam smearing, as a function of radius (blue data points), where each panel corresponds to the galaxy indicated in the legend. The black solid line in each panel is the median velocity dispersion in the disk of all points beyond a radius of 1.5$\times$ the beam FWHM. For comparison and as a check of our beam smearing correction, the \citet{girard21} molecular and ionized gas velocity dispersions and uncertainties are included as the black dashed, black dot dashed, and grey shaded regions respectively. The beam FWHM and channel width are represented by the error bars in the top left corner of each panel. 

\citet{girard21} obtained their velocity dispersion measurements from fitting the ALMA rotation curves using \textsc{GalPak3D} \citep[which corrects for beam smearing;][]{bouche15}, assuming a flat dispersion model. Beyond the central beam region of each galaxy, our beam smearing correction approach produces velocity dispersion results that are consistent with the results of \citet{girard21}: they are all within a channel width or less of each other.

The DYNAMO galaxies in our sample all appear to have higher velocity dispersions in the central beam region than they do in the disk. To assess the significance of this observation, we include as the solid yellow lines the median beam smearing correction, measured in annuli of increasing radius from our model dispersion maps (bottom row, middle panel of Figure \ref{fig:beam_smear}). Taking into account the channel width and comparing to the median beam smearing correction, it is possible to say that this may be the case for DYNAMO C13-1 and D13-5, and is very likely to be the case for DYNAMO G04-1, G08-5, and G14-1. These three galaxies are the ones for which we use the higher resolution ALMA observations. This, combined with the rotation curve turnover radius of ${<} 1$~kpc, suggests that the enhanced velocity dispersions we measure at smaller radii are likely real and not a consequence of beam smearing. 

\newpage
\bibliography{references}{}
\bibliographystyle{aasjournal}



\end{document}